\documentclass[10 pt]{article}
\usepackage{geometry}
\usepackage{amssymb}
\usepackage{amsmath}
\usepackage{graphicx}
\usepackage{bm,color}
\usepackage{epstopdf}
\begin{document}
\title{Second order gyrokinetic theory for Particle-In-Cell codes}

\author{Natalia Tronko$^{1}$, Alberto Bottino$^{1}$ and Eric Sonnendr\"{u}cker$^{1}$,
\\
$^{1}$Max-Planck-Institut f\"{u}r Plasmaphysik,  85748 Garching, Germany}

\maketitle

%
\begin{abstract}
%
The main idea of the gyrokinetic dynamical reduction consists in a systematical  removal of the fast scale motion (the gyromotion) from the dynamics of the plasma, resulting in a considerable simplification and a significant gain of computational time. The gyrokinetic Maxwell-Vlasov equations are nowadays implemented in for modeling (both laboratory and astrophysical) strongly magnetized plasmas.
Different versions of the reduced set of equations exist, depending on the construction of the gyrokinetic reduction procedure and the approximations performed in the derivation. The purpose of this article is to explicitly show the connection between the general second order gyrokinetic Maxwell-Vlasov system issued from the modern gyrokinetic theory and the model currently implemented in the global electromagnetic Particle-in-Cell code ORB5. Necessary information about the modern gyrokinetic formalism is given together with the consistent derivation of the gyrokinetic Maxwell-Vlasov equations from first principles.
The variational formulation of the dynamics is used to obtain the corresponding energy conservation law, which in turn, is used for the verification of energy conservation diagnostics currently implemented in ORB5.

This work fits within the context of the code verification project \textsf{VeriGyro} currently run at IPP Max Planck Institut in collaboration with others European institutions.
\end{abstract}
%
\section{\label{sec:intro}Introduction}
%

For more than five decades, magnetized plasmas have been investigated in order to achieve self-sustained nuclear reaction processes in fusion devices. Numerical simulations are necessary in order to better understand the dynamical behavior of plasmas. 
However these simulations rely on theoretical models compromising between an accurate description of the dynamics and a restricted number of numerical operations to keep simulations tractable in current computing facilities. 

Gyrokinetic theory aims at this compromise by taking advantage of the specific motion of the plasma. More precisely, the presence of a strong background magnetic field in such devices makes possible the separation of different scales. The main idea is to separate the fast motion of charged particles around the magnetic field lines (referred to as gyromotion) from their slower drift motion, in order to reduce the number of dynamical variables needed to describe the dynamics. 
The cyclotron frequency $\Omega=e B/m c$, where $e$ and $m$ are respectively the charge and mass of particles, $B$ is the magnetic field amplitude and $c$ is the speed of light, sets the scale of the gyromotion.
Mathematical tools and approximations allowing for the splitting out of this fast scale define a particular gyrokinetic dynamical reduction.

The gyromotion is described by a fast gyroangle variable $\theta$, to which corresponds a canonically conjugate slowly varying magnetic moment $\mu$, an adiabatic invariant of the system. At the lowest order of approximation $\mu = m v_\perp^2 /2 B$, where $v_\perp$ is the perpendicular velocity of particles respective to background magnetic field lines.
In early works \cite{Frieman_Chen_1982}, an iterative gyro-averaging procedure has been used in order to remove the $\theta$-dependence directly from the Vlasov equation. Such a procedure allowed for the derivation of non-linear gyrokinetic equations.
However the major issue was the impossibility to obtain an energy-conserving model from this procedure.

The modern gyrokinetic theory \cite{Brizard_Hahm,Littlejohn_1983} makes use of differential geometry (perturbative Lie-transformation techniques) to build up a new set of phase-space variables, such that the fast gyroangle variable $\theta$ becomes uncoupled from the description of particle's motion and the corresponding moment has trivial dynamics $\dot{\mu}=0$. Therefore, the particle phase space is reduced from $6$ dimensions to $4$+$1$ dimensions, which already represents a significant simplification for numerical simulations.

However one of the main difficulties is then to find a rigorous way to couple the reduced particle dynamics to those of the dynamical electromagnetic fields induced by the particles, in order to obtain a self-consistent description of the reduced dynamics. 

Two variational formulations exist, Lagrangian \cite{Low_1958,Sugama_2000} and Eulerian \cite{Cendra_Holm_Hoyle_Marsden_1998,brizard_prl_2000}, both providing a common framework for the description of gyrokinetically reduced self-consistent field-particles dynamics allowing for the derivation of energy and momentum conservation laws.
In these formulations, particles are described on the reduced phase space, while dynamical fields are still being evaluated at the non-reduced positions. One of the main advantages of the variational formulation is contained in the fact that polarization and magnetization effects arise naturally as a result of the dynamical reduction from the coupling with the dynamics of the reduced particles.

In the Lagrangian formulation, the dynamics of particles is represented by the characteristics, from which the Vlasov equation is reconstructed a posteriori. 
It allows one to choose a reduced model for the dynamics of the particles (e.g., linear polarization approach) and to systematically couple this reduced model to the dynamical electromagnetic fields.
However additional calculations are required to reconstruct the energy and momentum densities allowing for the derivation of conservation laws through Noether's theorem \cite{Scott_Smirnov}.
The Lagrangian formulation is the natural framework for Particle-in-Cell (PIC) code discretization \cite{Lee_1986}, \cite{Bottino_Sonnendrucker}.

Within the Eulerian approach, particles are represented by the Vlasov distribution function, which is treated as one of the dynamical fields of the theory. This leads to a direct derivation of conservation laws by Noether's method and does not require any external moments calculation \cite{Brizard_Tronko}.
This approach allows one to proceed with a systematic derivation of the reduced Maxwell-Vlasov model by truncating the action functional corresponding to the gyrokinetic system at the desirable order. 
At the same time, the Eulerian formulation is well suited to handle the splitting between the background and the fluctuating quantities. Such a manipulation on the Vlasov distribution function is used for the description of instabilities and could be particularly useful in order to keep ordering consistency within the reduced Vlasov equation.

In this article we derive a second order Maxwell-Vlasov gyrokinetic model from the systematic variational approach, suitable for code verification. We compare the result with the gyrokinetic equations recently implemented in the PIC code ORB5 \cite{Jolliet_2007}, \cite{Bottino_2011}.
We start our derivation by writing an explicit second order expression for the Eulerian action functional presented in Ref.~\cite{brizard_2010}. From the first variation of the action functional, we derive the corresponding Maxwell-Vlasov equations. We then derive the energy conservation law thanks to the Noether's method.
We finally get the Eulerian second order action principle corresponding to the gyrokinetic model implemented in ORB5 \cite{Jolliet_2007}.

This paper is organized as follows: in Section~\ref{sec:GK_particle_reduc} we explicit the necessary material for the derivation of the gyrokinetic reduction procedure for particle dynamics.
Section~\ref{sec:variational_general} starts with summarizing the main concepts of the Eulerian variational principle for the gyrokinetic Maxwell-Vlasov system.  We then derive the expression for the full second order Eulerian action and the corresponding reduced equations of motion. 

Section~\ref{sec:variational_explicit} starts with the derivation of the corresponding truncated Maxwell-Vlasov equations implemented in ORB5 from the corresponding Eulerian action functional. The obtained equations are then compared to the results obtained with the complete Eulerian variational principle. 

In Section \ref{sec:Energy_conservation_ORB5} Noether's method is used to get the expression for the conserved energy from the second order Eulerian action. Then we compare the result with the quantity implemented for energy conservation diagnostics in ORB5.

%
\section{\label{sec:GK_particle_reduc}Gyrokinetic dynamical reduction on particle's phase space: sources of polarisation and magnetisation.}
%
In this section we focus on the gyrokinetic dynamical reduction procedure for a single particle dynamics in external electromagnetic fields.
This is a preliminary step necessary for the derivation of the self-consistently reduced Maxwell-Vlasov model.
The main goal consists in exploring the intimate link between the reduced particle's dynamics and polarization/magnetization effects, which appear later on in the gyrokinetic Maxwell equations. 

Clarifying the effects of the particle dynamical reduction constitutes an important preliminary step for the correct coupling of the reduced dynamics with the electromagnetic fields and therefore for the gyrokinetic field theory.

The idea behind the gyrokinetic dynamical reduction is tightly related to the existence of an adiabatic invariant, the magnetic moment $\mu$, which in the simplest case of a slab magnetic geometry is given by  $\mu=m v_{\perp}^2/ 2 B$. The magnetic moment measures the area enclosed by the motion of a particle rotating around a magnetic field line. From this geometrical picture comes the idea of using $\mu$ as an action variable canonically conjugated to the fast gyromotion around magnetic field lines. 
In a straight uniform magnetic field, $\mu$ is an exact invariant. The effects of the  magnetic field curvature as well as the presence of fluctuating electromagnetic fields destroy this invariant. However, the average over long times of the magnetic moment is still being conserved $\langle\dot{\mu}\rangle_t=0$.

The dynamical reduction can be organized in one or two steps: In the one step case, the contributions from the background geometry non-uniformities and electromagnetic fluctuations to the breaking of the magnetic momentum conservation are taken into account simultaneously. In the two step case, these effects are treated in two separate stages.
Choosing the two step procedure is helpful for understanding the various contributions to the polarization and magnetization obtained from the gyrokinetic reduction. To each step corresponds a set of new phase space coordinates such that the fast gyromotion is uncoupled from the slow drifts of the particles. The dynamics on the reduced phase space is restricted to the surface $\dot{\mu}=0$.
These new coordinates are constructed as perturbative series of near-identity phase space transformations. These transformations are invertible at each step of the perturbative procedure.

Before we proceed with the detailed description of the reduction procedure, we first discuss the small parameters associated with each change of coordinates.
%
\subsection{\label{subsec:GK_orderings}Gyrokinetic orderings}
%

For the first step, called the guiding-center transformation, only the effects of the strong nonuniform background magnetic field are taken into account. We associate a small parameter $\epsilon_B=\rho_{th}/L_B$, representing the ratio between the thermal ion Larmor radius and the scale $L_B$ on which the background magnetic field exhibits important changes.  
We notice that in early works of Northrop and Littlejohn, the small parameter associated with the guiding-center dynamical reduction appears as a formal parameter, which scales as the inverse of the electric charge: $\epsilon\sim e^{-1}$.

For the second step, called the gyrocenter transformation, the reduced  guiding-center system is perturbed by external fluctuating electromagnetic fields. This leads to the mixing of time scales and therefore breaks down the conservation of the magnetic moment at the order of the amplitude of the perturbation. The goal of the  gyrocenter transformation is to restore the separation of time scales and the conservation of a slightly modified magnetic moment $\mu$ for the perturbed system. The small parameter related to that step of dynamical reduction measures the relative amplitude of the fluctuating fields $\epsilon_{\delta}=\epsilon_{\perp} e\delta\phi/T_i$, where $\epsilon_{\perp}=|{\mathbf k}_{\perp}\rho_{th}|$. Here $\delta\phi$ represents the amplitude of the fluctuating electrostatic potential and $T_i$ is the ion temperature.

For the gyrokinetic ordering consistency, one should consider the contributions from each dynamical reduction procedure at the same order: 
$\epsilon_B\sim\epsilon_{\delta}$.
However, in most of nowadays numerical simulations, the contributions from the background magnetic field curvature are pushed at the next order, i.e., 
$\epsilon_B\ll\epsilon_{\delta}$, which can be relevant for example for simulations with a large aspect ratio. 
%
\subsection{\label{subsec:L_particle}Gyrokinetic particle's Lagrangian}
%
In this section we proceed with the introduction of the central object of gyrokinetic dynamical reduction, the phase space  Lagrangian for a particle moving in external electromagnetic fields.
This time-dependent Lagrangian $L$ depending on the canonical variables $\left({\mathbf q},{\mathbf p},\dot{\mathbf q},\dot{\mathbf p}\right)$ writes:
\begin{eqnarray*}
L(\mathbf p, \mathbf q, \dot{\mathbf p}, \dot{\mathbf q},t)=\mathbf p\cdot \dot{\mathbf q}-\mathrm H(\mathbf{p},\mathbf{q},t).
\end{eqnarray*}
The choice of canonical coordinates is not always optimal. A classical example is problem of a charged particle moving in external electromagnetic fields. The canonical momentum mixes kinetic and space coordinates which leads to complications in the physical interpretation of the dynamics and the construction of the reduction procedure.

All kinds of invertible changes of variables are allowed and the dynamics is obtained from the Euler-Lagrange equations.
We consider an invertible change of variables ${\mathbf Z} ({\mathbf p}, {\mathbf q},t)$, such that the canonical coordinates are parametrized in the following way: $\mathbf p=\mathbf p\left({\mathbf Z},t\right)$ and ${\mathbf q}=\mathbf q\left(\mathbf Z,t\right)$. The phase space Lagrangian in the coordinates $\left(\mathbf Z,\dot{\mathbf Z},t\right)$ is given by:
\begin{eqnarray*}
L(\mathbf Z,\dot{\mathbf Z}, t)=p_i\frac{\partial q_i}{\partial Z_{\alpha}}\dot{Z}_{\alpha}+\mathbf{p}\cdot\frac{\partial{\mathbf q}}{\partial t}
-\mathrm{H}(\mathbf{p},\mathbf{q},t),
\end{eqnarray*}
where we define the coefficients of the symplectic part of the Lagrangian as 
$$
\Lambda_{\alpha}(\mathbf{Z},t)\equiv p_i\frac{\partial q_i}{\partial Z_{\alpha}},
$$ 
and the Hamiltonian 
$$ 
H(\mathbf Z,t) ={\mathrm H}(\mathbf{p},\mathbf{q},t)-\mathbf{p}\cdot\frac{\partial{\mathbf q}}{\partial t}.
$$
The phase-space Lagrangian $L(\mathbf Z,\dot{\mathbf Z}, t)$ now consists of a symplectic part $\bm{\Lambda}$ and a Hamiltonian part $H$:
\begin{equation}
L\left(\mathbf Z,\dot{\mathbf Z}, t\right)=\bm\Lambda(\mathbf{Z},t)\cdot\dot{\mathbf{Z}}-H(\mathbf Z,t)
\label{Lambda}
\end{equation}
The Euler-Lagrange equations in the new coordinates become
\begin{equation*}
\frac{d}{dt} \frac{\partial L}{\partial \dot{Z}_{\alpha}}=\frac{\partial L}{\partial Z_{\alpha}},
\end{equation*}
which can also be rewritten in the following form by using the explicit coordinate dependences of $\Lambda_{\alpha}$:
\begin{eqnarray*}
\left(\frac{\partial\Lambda_{\beta}}{\partial Z_{\alpha}}-\frac{\partial\Lambda_{\alpha}}{\partial Z_{\beta}}\right)\dot{Z}_{\beta}=
\frac{\partial H}{\partial Z_{\alpha}}+\frac{\partial\Lambda_{\alpha}}{\partial t}
\end{eqnarray*}
The components of the symplectic matrix 
$\omega$ are related to the components of the symplectic vector $\bm\Lambda$ as follows:
\begin{equation*}
\omega_{\alpha\beta}=\frac{\partial\Lambda^{\beta}}{\partial Z^{\alpha}}-\frac{\partial\Lambda^{\alpha}}{\partial Z^{\beta}},
\end{equation*}
which is related to the canonical coordinates in a following way through the  Lagrange bracket:
\begin{equation*}
\omega_{\alpha\beta}=\frac{\partial p_i}{\partial Z^{\alpha}}\frac{\partial q_i}{\partial Z^{\beta}}-
\frac{\partial q_i}{\partial Z^{\beta}}\frac{\partial p_i}{\partial Z^{\alpha}}\equiv \left[{Z}_{\beta},{Z}_{\alpha}\right]
\end{equation*}
In our problem $Z^{\alpha}$ represents the reduced phase space coordinates, which we will be explicitly defined in what follows.

Together the symplectic structure $\omega$ and the Hamiltonian $H$ provide us with the necessary information to derive the equations of motion on the reduced particle phase space.
In the case when the symplectic matrix is invertible we define the Poisson matrix as the inverse of the symplectic matrix $\Pi^{\alpha\beta}=\omega_{\alpha\beta}^{-1}$, and the Poisson bracket as $\left\{F,G\right\}=\frac{\partial F}{\partial Z^{\alpha}}\Pi^{\alpha\beta}\frac{\partial G}{\partial Z^{\beta}}$, we can write the reduced equations of motions 
$\dot{Z}^{\alpha}=\left\{Z^{\alpha},H\right\}=\Pi^{\alpha\beta}\frac{\partial H}{\partial Z^{\beta}}$.

In this work, we will specify the expression for the reduced Poisson bracket and the reduced Hamiltonian $H$ up to the second order in the dynamical reduction.

%
\subsubsection{\label{ssubsec: local_coord}Local particle's coordinates}
%

\begin{figure}[htbp]
\begin{center}
\includegraphics{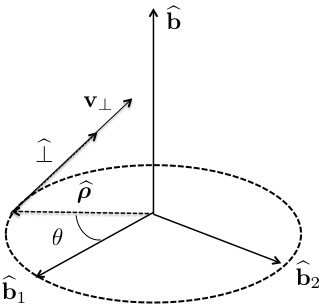}
\caption{The Frenet triad used for the definition of the local particle coordinates.}
\label{Frenet_triade}
\end{center}
\end{figure}

The dynamical reduction is performed in the \textit{local} particle coordinates $Z^{\alpha}$.  For this purpose, one needs to define two vector basis: the static one, which remains static as the particle rotates around the magnetic field line, and the dynamical one, which rotates with the particle.
As a static basis we consider the natural Frenet triad, associated with the unitary vector $\widehat{\mathbf b}= {\mathbf B}/B$, the direction of the background magnetic field. In the perpendicular to the background magnetic field plane we use the normalized curvature vector $\widehat{\mathbf b}_1= \widehat{\mathbf b}\cdot\bm\nabla\widehat{\mathbf b}/\left|\widehat{\mathbf b}\cdot\bm\nabla\widehat{\mathbf b}\right|$ and we define the third basis vector as $\widehat{\mathbf b}_2=\widehat{\mathbf b}\times\widehat{\mathbf b}_1$.
The dynamical triad is constructed from the static one as follows (see also Fig.~\ref{Frenet_triade}): We take $\widehat{\mathbf b}$ as its first vector and we define
\begin{eqnarray}
\widehat{\bm{\rho}}=\widehat{\mathbf b}_1\cos\theta -\widehat{\mathbf b}_2\sin\theta,
\label{rho_hat}
\end{eqnarray}
and 
\begin{eqnarray}
\widehat{{\perp}}=-\widehat{\mathbf b}_1\sin\theta -\widehat{\mathbf b}_2\cos\theta,
\label{perp_hat}
\end{eqnarray} 
where $\theta$ is the gyrophase angle defined according to the direction of the perpendicular velocity in the following way: 
\begin{equation*}
\mathbf{v}=v_{\|}\widehat{\mathbf b}+\mathbf v_{\perp} \equiv v_{\|}\widehat{\mathbf b}+ \sqrt{\frac{2\mu B}{m}}\widehat{\perp},
\end{equation*}
and $\mu=\frac{mv_{\perp}^2}{2 B}$.
The new local particle coordinates are now ${\bf Z}=(\mathbf x, v_{\|}, \mu, \theta)$.
%
\subsubsection{\label{ssubsec:purpose_gk_red}Purpose of gyrokinetic dynamical reduction}
%
The goal of the dynamical reduction procedure is to define a near-identity change of coordinates on particle phase space, such that the fast dynamics associated with the gyromotion is uncoupled from the dynamical description. The exact conservation of the adiabatic invariant $\mu$ is considered as a constraint at each order of the iterative procedure. The coordinates in the particle phase space are related to the coordinates in the reduced phase space by a series of Lie transforms: $Z^{\alpha}=\exp(\pounds_{{S}_n})\overline{Z}^{\alpha}$, where ${ S}_n$ is a generator of a near-identity transformation at the order $\mathcal{O}(\epsilon^n)$ defined from the general perturbative procedure\cite{Tronko_Brizard_2015}. 
It is important to notice that in the general case, this operator $\pounds_{{S}_n}$ is gyrophase, i.e., $\theta$-dependent. In other words, it means that in order to rewrite the dynamics in the new variables, which allow the uncoupling of the fast dynamics, one needs to perform a coordinate transformation depending itself on this fast variable.

An important element of the gyrokinetic dynamical reduction is that the dynamics is changed in the Hamiltonian $\overline{H}=\exp(\pounds_{{ S}_n})H$ while the Poisson bracket remains unchanged under the Lie transform.
From the point of view of the reduced dynamics, at each order of the gyrokinetic coordinate transformation the gyrophase dependence is pushed to the next order.

%
\subsection{\label{subsec:guiding_center}Dynamical reduction, first step: guiding-center dynamics}
%
We are now proceeding with a detailed description of the dynamical reduction procedure. The first step considers the effects of the background magnetic field only on the guiding-center dynamics. 
The guiding-center dynamical reduction is time-independent. This is why it is performed on a $6$ dimensional phase space, consisting of the parallel kinetic momentum $p_{\|}=m v_{\|}$, the reduced guiding-center position $\mathbf X$, the magnetic moment $\mu$ and the gyroangle $\theta$.
The corresponding Lagrangian writes
\begin{equation}
L_{\mathrm{gc}}\left(\mathbf{X}, p_{\|},\mu,\theta\right) =\frac{e}{c}\mathbf {A}^*\cdot\dot{\mathbf X}+\frac{m c}{e}\mu\ \dot{\theta}-H_{\mathrm{gc}},
\label{gc_Lagrangian}
\end{equation}
where the symplectic part contains the modified magnetic potential:
\begin{equation}
{\mathbf A}^*=\mathbf{A}+\frac{c}{e}\ p_{\|}\widehat{\mathbf b}.
\label{A*}
\end{equation}
The guiding-center Hamiltonian is given by:
\begin{equation}
H_{\mathrm{gc}}=\frac{p_{\|}^2}{2 m}+\mu B.
\label{H_gc}
\end{equation}
By inverting the symplectic matrix which corresponds to the Lagrangian (\ref{gc_Lagrangian}) we obtain the following guiding-center Poisson bracket:
\begin{eqnarray}
\{F,G\}_{\mathrm{gc}}=\frac{e}{mc} \left(\frac{\partial F}{\partial \theta}\frac{\partial G}{\partial \mu}-\frac{\partial F}{\partial \mu}\frac{\partial G}{\partial \theta}\right)+\frac{{\mathbf B}^*}{B_{\|}^*}\cdot\left(\bm\nabla F\frac{\partial G}{\partial p_{\|}}-\frac{\partial F}{\partial p_{\|}}\bm\nabla G\right)-
\frac{c\widehat{\mathbf b}}{e B_{\|}^*}\cdot\left(\bm\nabla F\times\bm\nabla G\right),
\label{GC_PB}
\end{eqnarray}
where ${\mathbf B}^*\equiv \nabla\times{\mathbf A}^*$ represents the modified magnetic field.
We notice that the condition $\bm\nabla\cdot{\mathbf B^*}=0$ guarantees the Liouville theorem (i.e., the conservation of the phase space volume) on the reduced phase space (see Ref.~\cite{Brizard_Hahm} for more details). 

At the same time, the condition of the invertibility of the Lagrange matrix is equivalent to $B_{\|}^*\neq0$.
In the case where this condition is not fulfilled, the gyrokinetic dynamical reduction cannot be performed. From the physical point of view, it means that the amplitude of the background magnetic field $B$ is comparable to $c/e p_{\|}\bm\nabla\times\widehat{\mathbf b}$. Taking into the account that the curvature terms are small: $\left|\bm\nabla\times\widehat{\bf b}\right|\sim\mathcal{O}(\epsilon_{B})$ and the background magnetic field is strong, it can only happen if the value of the parallel kinetic momentum $p_{\|}$ becomes very large. This situation is outside the range of applicability of the gyrokinetic theory.

The characteristics of the guiding-center dynamics are derived as follows:
\begin{eqnarray}
\dot{\mathbf X}&=&\left\{ {\mathbf X}, H_{\mathrm{gc}}\right\}_{\mathrm{gc}}= \frac{p_{\|}}{m}\ \frac{\mathbf{B}^*}{B_{\|}^*} +\frac{c\widehat{\mathbf b}}{e B_{\|}^*}\times
\mu\bm\nabla B,
\label{X_gc}
\\
\dot{p_{\|}}&=&\left\{p_{\|}, H_{\mathrm{gc}}\right\}_{\mathrm{gc}}=-\mu\bm\nabla B \cdot\frac{\mathbf{B}^*}{B_{\|}^*} ,
\label{p_par_gc}
\\
\dot{\theta}&=&\frac{e B}{m c},
\label{theta_gc}
\\
\dot{\mu}&=&0
\label{mu_gc}.
\end{eqnarray}

The fastest scale of motion $(\ref{theta_gc})$ is uncoupled from the reduced position (\ref{X_gc}) and the kinetic momentum dynamics (\ref{p_par_gc}). At the same time, the magnetic moment $\mu$ has a trivial dynamics on the reduced phase space. 
%
\subsection{\label{subsec:gyrocenter}Beyond the guiding-center reduction: gyrocenter dynamics}
%
In the framework of the two-step gyrokinetic reduction, external time-dependent electromagnetic fields are introduced into the system at the second step of the dynamical reduction: the \textit{gyrocenter} step. 
The guiding-center dynamics is perturbed by the electromagnetic potentials
$\phi_{1\mathrm{gc}}$ and $A_{1\|\mathrm{gc}}$ evaluated at the guiding-center position $\mathbf X$.
The perturbed phase space Lagrangian becomes
\begin{eqnarray}
\widetilde{L}_{\mathrm{gc}}\left(\mathbf X, p_\parallel, \mu, \theta; t \right)=\left(\frac{e}{c}\mathbf{A}+\frac{e}{c}A_{1\|\mathrm{gc}}\widehat{\mathbf b}+p_{\|}\widehat{\mathbf b}\right)\cdot \dot{\mathbf X}
+\frac{mc}{e}\mu\ \dot{\theta}-\left(\frac{p_{\|}^2}{2m}+\mu B +e \phi_{1\mathrm{gc}}\right).
\label{L_gc_pert_gen}
\end{eqnarray}
By performing the following change of momentum,
\begin{equation}
\label{eqn:pz}
p_z=p_{\|}+\frac{e}{c} A_{1\| \mathrm{gc}},
\end{equation}
we transfer the perturbation from the symplectic part to the Hamiltonian one:
\begin{eqnarray}
\widetilde{L}_{\mathrm{gc}}(\mathbf X, p_{z},\mu,\theta; t)&=&\left(\frac{e}{c}{\mathbf A}+p_z\widehat{\mathbf b}\right)\cdot \dot{\mathbf{X}} +\frac{mc}{e}\mu\ \dot{\theta} 
\label{L_gy_pert}
\\
&-&\left(\frac{p_z^2}{2m}+\mu B
+e\phi_{1\mathrm{gc}}-\frac{e p_z}{m c}A_{1\|{\mathrm{gc}}}
+\frac{1}{2m}\left(\frac{e}{c}\right)^2 A_{1\| \mathrm{gc}}^2\right).
\nonumber
\end{eqnarray}
Due to the time-dependence of the fluctuating fields, time becomes a dynamical variable of the system. Therefore, one needs to extend the guiding-center phase space by introducing a couple of canonically conjugated variables $(w,t)$.
The extended non-perturbed  Hamiltonian is ${\mathcal H}_{\mathrm{gc}}\equiv H_{\mathrm{gc}}-w$ and  the Poisson bracket (\ref{GC_PB}) has an additional term:
\begin{eqnarray}
\{F,G\}_{\mathrm{ext}}&=&\frac{e}{mc} \left(\frac{\partial F}{\partial \theta}\frac{\partial G}{\partial \mu}-\frac{\partial F}{\partial \mu}\frac{\partial G}{\partial \theta}\right)+\frac{{\mathbf B}^*}{B_{\|}^*}\cdot\left(\bm\nabla F\frac{\partial G}{\partial p_{\|}}-\frac{\partial F}{\partial p_{\|}}\bm\nabla G\right)
\nonumber
\\&-&
\frac{c\widehat{\mathbf b}}{e B_{\|}^*}\cdot\left(\bm\nabla F\times\bm\nabla G\right)
-\frac{\partial F}{\partial w} \frac{\partial G}{\partial t}+\frac{\partial F}{\partial t}\frac{\partial G}{\partial w}.
\label{GC_PB_EXT}
\end{eqnarray}
Doing so, we keep the symplectic part of Eq.~(\ref{L_gy_pert}) unchanged and therefore the gyrocenter Poisson bracket does not depend on the fluctuating fields. This is the common choice, which has been already adopted in the derivation of the model in ORB5. All the effects from the dynamical reduction must be accounted inside the expression for the reduced gyrocenter Hamiltonian $H_{\mathrm{gy}}$.

Before proceeding with the derivation of this reduced Hamiltonian, we first discuss the consequence of using a new reduced particle position, the gyrocenter position, on the polarization corrections.
%
\subsubsection{\label{ssubsec:polariz_origin}Polarization effects: relationship between the coordinate change and the reduced Hamiltonian dynamics}
%
The systematic reduction procedure applied to the particle phase space Lagrangian provides a set of new coordinates, in which the reduced particle dynamics is described.
Following the general reduction procedure~\cite{Brizard_Hahm,Tronko_Brizard_2015}, we would like to stress that the definition of some new coordinates is intimately related to the identification of polarization corrections due to the dynamical reduction into the reduced Hamiltonian. We adopt the general strategy used in numerical codes, namely, we push the curvature effects of the magnetic field at the next order with respect to the amplitude of the electromagnetic fluctuations: $\epsilon_B=\epsilon\ \epsilon_{\delta}$, where $\epsilon<1$ is a free small parameter, which can also depend on $\epsilon_{\delta}$. Of course, different choices of $\epsilon$ will lead to different models. Here we consider our series expansions up to the second order in $\epsilon_{\delta}$ and the first order in $\epsilon_{B}$, which lead to the same result as with $\epsilon_B=\epsilon_{\delta}^2$ or $\epsilon_B=\epsilon\ \epsilon_{\delta}$. We notice that the norm of $\bm\rho_0$ is of order $\mathcal O (\epsilon_B^0)$ and the one of $(\bm\rho_0\cdot\bm\nabla)\bm\rho_0$ is of order $\mathcal O (\epsilon_B)$.

Here we focus on the explicit derivation of the reduced Hamiltonian, by assuming that we have performed the dynamical reduction at the lowest order of the guiding-center and the gyrocenter  transformations. 
In the other words, it means that the difference between the initially non-reduced particle position $\mathbf x$ and the reduced position $\mathbf X$ is defined as follows:
\begin{eqnarray}
\mathbf x\equiv\mathbf X+\bm\rho_0(\mathbf X,\mu,\theta)+\bm\rho_1(\mathbf X,\mu,\theta),
\label{coord_transform}
\end{eqnarray} 
with $\bm{\rho}_0$ corresponding to the lowest order guiding-center displacement and $\bm\rho_1$ the lowest order of the gyrocenter displacement.
The lowest order guiding-center displacement is given by:
\begin{equation}
\bm\rho_{0}\equiv\frac{m c }{e}\sqrt{\frac{2 \mu}{m B}}\ \widehat{\bm\rho}\equiv\rho_0\widehat{\bm\rho},
\label{rho_0}
\end{equation}
where $\widehat{\bm\rho}$ is one of the dynamical basis vectors defined in Eq.~(\ref{rho_hat}).
This displacement takes into account the background magnetic field $\mathbf B$, which is locally uniform. We emphasize that the amplitude of the background magnetic field $B$ is evaluated at the reduced guiding-center position $\mathbf X$. All the following corrections to the guiding-center displacement are related to the magnetic curvature.
The expression for the first order gyrocenter displacement is given by:
\begin{equation}
\bm\rho_1\equiv -\frac{m c^2}{B^2} \bm\nabla_{\perp}\left( \phi_1(\mathbf X)-\frac{p_z}{mc}A_{1\|}(\mathbf X) \right).
\label{rho_1}
\end{equation}
The detailed derivation of the expression for $\bm\rho_1$ is presented in Appendix \ref{app: rho1}.

Deriving the reduced expression for the reduced Hamiltonian is crucial for the derivation of the reduced Maxwell-Vlasov equations. It allow us to define polarization effects, i.e., effects due to the dynamical reduction in the reduced Maxwell equations later on from the variational principle.
In particular, we show that for the second order corrections in $\epsilon_{\delta}$ to the reduced Hamiltonian requires a coordinate transformation of the first order in $\epsilon_{\delta}$.

%
\subsubsection{\label{ssubsec:H2}Second order reduced Hamiltonian}
%
We start by constructing the perturbed electromagnetic Hamiltonian ${H}_{\mathrm{gy}}$ up to the second order in $\epsilon_{\delta}$.
First, we need to specify the approximations performed on electromagnetic fields. We consider the low frequency approximation for the electric field: ${\mathbf E}_{1\perp}\equiv-\bm\nabla_{\perp}\phi_1$ for the perpendicular component, and $E_{1\|}=-\widehat{\mathbf b}\cdot\bm{\nabla}\phi_1-c^{-1}\partial_t A_{1\|}$ for the parallel one. Therefore, the parallel component of the electric field is of the next order with respect to the gyrocenter parameter $\epsilon_{\delta}$ than the perpendicular one: 
$|E_{1\|}|\sim\epsilon_{\delta} |{\mathbf E}_{1\perp}|$. The magnetic perturbation is given by:
\begin{equation}
{\mathbf B}_1=\bm\nabla \times (\widehat{\mathbf b} \ A_{1 \|}),
\label{B_1_par}
\end{equation}
i.e., the perturbed magnetic potential has only a parallel component.
It satisfies $\bm\nabla\cdot{\mathbf B}_1=0$ and therefore the Liouville theorem for the phase space volume conservation on the reduced phase space applies. This perturbation leads to the appearance of magnetic curvature terms in the Amp\`ere's law and the conservation laws.
Later in Sec.~\ref{sec:variational_explicit} we remark that, taking into account only the parallel fluctuations of the perturbed magnetic potential leads to the derivation of only the parallel component of Amp\`ere's law from the variational principle.

Following the $p_z$-representation of the Lagrangian defined in Eq.~(\ref{L_gy_pert}),
the reduced gyrocenter Hamiltonian is:
\begin{equation}
H_{\mathrm{gy}}=\frac{p_{z}^2}{2 m}+\mu B+e\ \psi_{1 \mathrm{gc}}({\mathbf{X}})
+\frac{1}{2 m}\left(\frac{e}{c}\right)^2 A_{1\|}({\mathbf{X}}+\bm\rho_0)^2,
\label{eqn:Hgy}
\end{equation}
where the linear perturbed gyrocenter potential is:
\begin{equation}
\psi_{\mathrm{1 gc}}({\bf X})\equiv\ \phi_{1}({\mathbf{X}}+\bm\rho_0)
-\frac{1}{m c}\ p_{z} \ A_{1\|}({\mathbf{X}}+\bm\rho_0).
\label{psi_1}
\end{equation}

In order to eliminate the gyrophase dependence of Hamiltonian $H_{\mathrm{gy}}$, we perform a Lie transform which maps the guiding-center coordinates into the gyrocenter ones. The Hamiltonian expressed in these new coordinates, up to the second order in $\epsilon_{\delta}$ is
\begin{equation}
\overline{H}_{\mathrm{gy}}
=e^{-\pounds_{S_2}} e^{-\pounds_{S_1}}H_{\mathrm{gy}},
\end{equation}
where $S_1$ and $S_2$ are the generating functions, defined in order to remove the gyrophase dependence at the orders $\epsilon_{\delta}$ and  $\epsilon_{\delta}^2$ respectively. The derivation of Hamiltonian $\overline{H}_{\mathrm{gy}}$ is detailed in Appendix~\ref{app: rho1}, and it leads to the following expression:
\begin{eqnarray}
\overline{H}_{\mathrm{gy}}
&=&
\frac{p_z^2}{2 m} +\mu B+ \epsilon_{\delta}\left(e \left\langle \phi_1 (\mathbf X+\bm\rho_0)\right\rangle
-\frac{e}{mc}\ p_z\ \left\langle A_{1\|} (\mathbf X+\bm\rho_0)\right\rangle\right)
\label{H_1gy}
\\
&+&\epsilon_{\delta}^2\left( 
\frac{1}{2 m} \left(\frac{e}{c}\right)^2 \left\langle A_{1\|}(\mathbf X+\bm\rho_0)^2\right\rangle
-
\frac{m c^2}{2 B^2}\left| \bm\nabla_{\perp} \phi_1(\mathbf X)-\frac{e}{c}\ p_z\bm\nabla_{\perp}A_{1 \|} (\mathbf X) \right|^2\right),
\nonumber
\end{eqnarray}
where we have explicitly introduced the parameter $\epsilon_\delta$ for bookkeeping purposes, so as to clearly identify which terms are of order $\epsilon_\delta$ and which are of order $\epsilon_\delta^2$. The average over the gyroangle is given by 
$$
\left\langle F \right\rangle=\frac{1}{2\pi}\int_0^{2\pi}\ d\theta\ F({\bf X},\mu,\theta). 
$$

Here we introduce the models for the particle dynamics, which we use in Sec.~\ref{sec:variational_explicit} for the variational derivation of the gyrokinetic Maxwell-Vlasov equations.
The first order (with respect to $\epsilon_{\delta}$) correction $H_1$ to the general gyrocenter Hamiltonian  (\ref{H_1gy}) is given by:
\begin{eqnarray}
H_1&=&\epsilon_{\delta}\ e\ \psi_{1\mathrm{gc}}(\bf X),
\label{H1}
\end{eqnarray}
where $\psi_{1\mathrm{gc}}(\bf X)$ is defined in Eq.~(\ref{psi_1}).

Concerning the second order Hamiltonian $H_2$, we truncate the squared parallel magnetic potential to the second order in the guiding-center FLR corrections:
\begin{equation}
A_{1\| \mathrm{gc}}=A_{1\| } ({\mathbf X}) +{\bm\rho}_{0}\cdot\bm\nabla A_{1\|}(\mathbf X)+\frac{1}{2}{\bm\rho}_{0}{\bm\rho}_{0} :
\bm\nabla\bm\nabla A_{1\|}(\mathbf X).
\label{Second_FLR_A_1}
\end{equation}
With using the expression for the gyro-averaged dyadic tensor $\widehat{\bm\rho}\widehat{\bm\rho}$, we write the explicit expression for the second order Hamiltonian $H_2$ with FLR effects up to the second order: 
\begin{eqnarray}
H_2=
\frac{e^2}{2 m c^2} A_{1\|}^2+\frac{\mu}{2 B} \left|\bm\nabla_{\perp}A_{1\|}\right|^2
+\frac{1}{2}\frac{\mu}{B}A_{1\|}\bm\nabla_{\perp}^2 A_{1\|}
-
\frac{m c^2}{2 B^2} \left|\bm\nabla_{\perp}\phi_{1}-\frac{p_{z}}{m c}\bm\nabla_{\perp}A_{1\|}\right|^2. 
\label{H2_full_GK}
\end{eqnarray}
The first and the second terms of Eq.~(\ref{H2_full_GK}) are the first order guiding-center FLR corrections to the averaged gyrocenter magnetic potential, and the third term is the second order guiding-center FLR contribution. The last term represents the lowest order gyrocenter polarization correction associated with the gyrocenter displacement $\bm\rho_1$ given by Eq.~(\ref{rho_1}). 
As it can be seen, the latter is related to the gradient of the electromagnetic potential in the general gyrokinetic theory. 
However, we notice that the model for ORB5 as well as most physical models consider its electrostatic part only. 
\subsubsection{\label{ssubsec:Hamilt_Symplectic}Symplectic and Hamiltonian representations of the gyrocenter reduction}
As we have seen in the previous subsection, there exist two possibilities for writing down the perturbed phase-space Lagrangian (\ref{L_gc_pert_gen}) depending on where the magnetic potential perturbation $A_{1\|}$ is taken into account, into the Symplectic or into the Hamiltonian part.

The Hamiltonian representation includes the parallel magnetic perturbation $A_{1\|}$ in the expression for the perturbed Hamiltonian and leaves the guiding-center Poisson bracket (\ref{GC_PB_EXT}) unchanged. As we have seen in the previous section, the parallel canonical gyrocenter momentum $p_z$ is used as one of the phase space variables. It is therefore sometimes called the "$p_z$ representation".
 
The Symplectic representation accounts for the perturbed parallel magnetic moment in the Symplectic part of the phase-space Lagrangian (\ref{L_gc_pert_gen}), which leads to a different Poisson bracket than the one given in Eq.~(\ref{GC_PB_EXT}). In this case, the parallel kinetic momentum $p_{\|}=m v_{\|}$ is kept. This representation facilitates the identification of various physical terms and avoids the cancellation problem related to presence of terms with different orders in the corresponding Amp\`ere's equation \cite{Hatzky_2007}.

In what concerns PIC codes, the Hamiltonian representation is preferable, since in the symplectic one the inductive electric field (the explicit time-derivative of the perturbative magnetic potential) appears in the characteristics. It therefore requires an explicit time integrator. In this case, the explicit time derivative of the perturbed potentials is only contained in the dynamics of the variable $w$ and therefore completely uncoupled from the dynamics of the physical reduced phase space. The Symplectic representation is used for the derivation of the particle characteristics and the Vlasov equation implemented in the Eulerian framework, e.g., GENE code \cite{Jenko_2000,Goerler_2011}.

\paragraph*{First order gyrocenter characteristics}
At the first order in $\epsilon_\delta$ the particle characteristics in the $p_z$ representation are derived from the first order gyrocenter Hamiltonian:
\begin{equation}
{\mathcal H}_{\mathrm{gy}}^{(1)}=H_{\mathrm{gc}} + e\ \epsilon_{\delta} \left( \left\langle\phi_{\mathrm{1 gc}}\right\rangle-\frac{1}{mc} \ p_{z}\left\langle A_{1\| \mathrm{gc}}\right\rangle\right) -w,
\label{H_gy_ext}
\end{equation}
and using the non-perturbed guiding-center Poisson bracket on the extended $8$-dimensional phase space (\ref{GC_PB_EXT}), we obtain
\begin{eqnarray}
\dot{\mathbf X} &=&\left\{ {\mathbf X}, {\mathcal H}_{\mathrm{gy}}^{(1)}\right\}_{\mathrm{ext}}=\frac{{\mathbf B}^*}{B_{\|}^*}\frac{\partial {\mathcal H}_{\mathrm{gy}}^{(1)}}{\partial p_{z}} +
\frac{c\hat{\mathbf b}}{e B_{\|}^*}\times\bm\nabla {\mathcal H}_{\mathrm{gy}}^{(1)},
\label{GY_CHAR_1}\\
\dot{p}_{z} &=&\left\{p_{z}, {\mathcal H}_{\mathrm{gy}}^{(1)}\right\}_{\mathrm{ext}}=-\frac{{\mathbf B}^*}{B_{\|}^*}\cdot\bm\nabla {\mathcal H}_{\mathrm{gy}}^{(1)}.
\label{GY_PZ_1}
\end{eqnarray}

In Appendix \ref{app:ORB5_diagnostics}, we give a detailed derivation of the characteristics equations in the Hamiltonian representation as it is implemented in ORB5, together with the ORB5 code diagnostics.

In the Symplectic representation, only the electrostatic part of the perturbation is included inside the expression for the gyrocenter perturbed Hamiltonian 
$\mathcal H_{\mathrm{gy}}^s=H_{\rm gc}+e\epsilon_\delta \langle \phi_{1 \rm gc}\rangle-w$, while the magnetic part of the perturbation is taken into account into the symplectic part of the Lagrangian (\ref{Lambda}) through the symplectic potential (\ref{A*}) and therefore into the modified Poisson bracket $ \{\cdot ,\cdot\}_{\rm gy}$ (see Ref.~\cite{Brizard_Hahm} for more details). This representation contains explicitly the time derivatives of the magnetic potential in the expressions for the reduced phase space characteristics $({\mathbf X},p_{\|})$:
\begin{eqnarray*}
\dot{\mathbf X}=\left\{ {\mathbf X}, {\mathcal H}_{\mathrm{gy}}^s\right\}_{\mathrm{gy}}=\frac{{\mathbf B}^*}{B_{\|}^*}\frac{\partial {\mathcal H}_{\mathrm{gy}}^s}{\partial p_{\|}} +
\frac{c\hat{\mathbf b}}{e B_{\|}^*}\times\bm\nabla {\mathcal H}_{\mathrm{gy}}^s\\
\dot{p_{\|}}=\left\{p_{\|}, {\mathcal H}_{\mathrm{gy}}^s\right\}_{\mathrm{gy}}=-\frac{{\mathbf B}^*}{B_{\|}^*}\cdot\bm\nabla {\mathcal H}_{\mathrm{gy}}^s-\frac{e}{c}\frac{\partial \left\langle A_{1\| gc}\right\rangle}{\partial t},
\end{eqnarray*}
where the symplectic magnetic field $\mathbf B^*$ now contains the perturbed magnetic field.
%
\section{\label{sec:variational_general}Eulerian second order variational principle: general method}
%

In this section we consider an expression for the gyrokinetic Eulerian action functional up to order
$\mathcal O(\epsilon_{\delta}^2)$ and up to the second order in the guiding-center Finite Larmor Radius (FLR) effects related to the lowest order displacement $\bm\rho_0$. The expression of the Eulerian action functional has been obtained in Ref.~\cite{brizard_2010} by a systematic truncation of the full Eulerian gyrokinetic action functional \cite{brizard_prl_2000}. 
It writes:
\begin{eqnarray}
{\mathcal I}^{\mathcal E} [{\phi}_1,{\mathbf A}_{1}, {\mathcal F}]&\equiv&
 \int_{t_1}^{t_2}{\mathcal A}^{\mathcal E} [{\phi}_1,{\mathbf A}_{1}, {\mathcal F}]\ dt
 =\int \frac{dV\ dt\ }{8 \pi} \left(\epsilon_{\delta}^2 \left|{\mathbf E}_1 \right|^2-
\left|\mathbf{B}_0+\epsilon_{\delta}\mathbf{B}_1\right|^2\right) 
\nonumber
\\
&-&
\int\ \ d^8 Z \ \mathcal{F}\ {\mathcal H}_{1}
\label{Euler_2_action}
-\epsilon_{\delta}^2\int\ d^6 Z\ dt  \ F_0\ H_2 ,
\end{eqnarray}
where ${\bf E}_1=-\bm\nabla \phi_1-c^{-1}\partial_t {\bf A}_1$ and ${\bf B}_1=\bm\nabla \times {\bf A}_1$ and
\begin{equation}
\mathcal{F}\equiv\left(F_0+\epsilon_{\delta} \ F_1\right)\delta\left(w-H_{0}- \epsilon_{\delta} H_1\right) , 
\label{F_tr_ext}
\end{equation}
and 
$$
{\cal H}_1\equiv H_0+\epsilon_\delta H_1 -w,
$$
in the extended phase space with variables ${\bf Z}=({\bf X},t,\mu, p_{z},\theta,w)$. The measure element is $d^8 Z \equiv\ B_{\|}^*d V\ d p_{z}\  d \mu\ d\theta\ d t\ d w$ where $dV=d^3X$. 
The first term of Eq. (\ref{Euler_2_action}) represents the Maxwell's part of the action with electrostatic field ${\mathbf E}_1\equiv -\bm{\nabla} \phi_1-c^{-1}\partial_t {\mathbf A}_1$. 
The magnetic field is separated into a background ${\mathbf B}_0$ and a fluctuating (dynamical) part ${\mathbf B}_1\equiv\bm{\nabla}\times{\mathbf A}_1$.
The second and third terms are contributions to the Vlasov part of the action. The first Vlasov term contains dynamical part of the distribution function $F_1$ defined on the extended $8$-dimensional phase space, and $F_0$ is the non-dynamical (background) part of the Vlasov distribution. The second Vlasov term is defined on the $6$-dimensional reduced gyrocenter phase space with $d^6 Z \equiv\ B_{\|}^*d V\ d p_{z}\  d \mu\ d\theta$ and contains the second order reduced gyrocenter dynamics, generated by Hamiltonian $H_2$, and therefore, associated with the background distribution function $F_0$ only.

Here our goal is to compare the results of our derivation (i.e., equations of motion and conservation laws) for the Eulerian action functional (\ref{Euler_2_action}) with the action functional used for construction of the ORB5 code model. We describe this model in Section \ref{sec:ORB5_variational}.
Finally, we aim to build up an exactly conserved electromagnetic energy invariant via Noether's method. We discuss this issue in Section \ref{subsec:Noether_method}. Here we proceed with the derivation without using the explicit expressions for $H_1$ and $H_2$ in order to simplify the derivation. 
%
\subsection{\label{subsec:implicit}First variation of the action functional}
%
\label{sec:IIIA}
Here we briefly describe the procedure for the first variation calculation of the first variation of the second order Eulerian action (\ref{Euler_2_action}). 

As a reminder, for a functional $\mathcal{L}=\int d\Lambda \ {\mathsf L}\left(\eta,\nabla\eta\right)$ depending on a scalar field $\eta=\eta({\mathbf x})$ and its gradient $\bm \nabla\eta=\bm \nabla\eta({\mathbf x})$,
the functional derivative is defined as:
\begin{eqnarray}
\frac{\delta{\mathcal L}}{\delta\eta}\circ\chi\equiv
\left.\frac{d}{d\nu} \left[\int d\Lambda\ {\mathsf L}\left(\eta+\nu\chi,\bm\nabla\eta+\nu\bm\nabla\chi\right)\right]\right|_{\nu=0}=
\int d\Lambda\ \frac{\partial{\mathsf L}}{\partial\eta}\circ\chi+\int d\Lambda\ \frac{\partial{\mathsf L}}{\partial\bm\nabla\eta}\circ\bm\nabla\chi
\label{weak_eqns}
\\
=\int d\Lambda\ \left(\frac{\partial{\mathsf L}}{\partial{\eta}}-
\left(\bm\nabla\cdot\frac{\partial{\mathsf L}}{\partial\bm\nabla\eta}\right)\right)\circ\chi
+\int d\Lambda\ \bm\nabla\cdot\left(\frac{\partial{\mathsf L}}{\partial\bm\nabla\eta}\circ\chi\right),
\label{func_derivative}
\end{eqnarray}
where $\chi$ is an arbitrary test function. We call the first term of Eq.~(\ref{func_derivative}) the \textit{dynamical} term and the second term, the \textit{Noether} term. The first variation for the second order Eulerian action (\ref{Euler_2_action}) is:
\begin{eqnarray}
\delta{\mathcal I}^{\mathcal E}\equiv \int_{t_1}^{t_2} \delta\mathcal{A}^{\mathcal E} \left[\phi_1,{\mathbf A}_1,{\mathcal F}\right] dt
=
\int_{t_1}^{t_2}\left(
\frac{\delta{\mathcal A}^{\mathcal E}}{\delta\phi_1}\circ\widehat{\phi}_1+
\frac{\delta{\mathcal A}^{\mathcal E}}{\delta \mathbf{A}_{1}}\circ\widehat{\mathbf A}_{1}+
\frac{\delta{\mathcal A}^{\mathcal E}}{\delta {\mathcal F}}\circ\delta \widehat{\mathcal F}
\right) dt,
\label{first_var_implicit_amp}
\end{eqnarray}
where $\widehat{\phi}_1$ and $\widehat{\mathbf A}_1$ are test functions and $\delta \widehat{\cal F}$ is a constrained variation to be specified later.

Such a derivation provides two important informations: First of all, it allows one to get a system of coupled Maxwell-Vlasov equations. Second, it also provides the expressions for the Noether terms necessary for the derivation of the corresponding conservation laws. The Noether terms are 
represented by exact derivatives and do not contribute to the dynamical part.

To that purpose, while evaluating the first variation of the action functional ${\mathcal I}^{\mathcal E}[\phi_1,\mathbf{A}_{1},{\mathcal F}]$, we are using the above definition (\ref{func_derivative}) of the functional derivative which makes the dependences on the scalar fields and their gradients explicitly appear .
As we will see, Eq.~(\ref{weak_eqns}) defines the dynamics in a \textit{weak} form,
while Eq.~(\ref{func_derivative}) obtained by applying the Leibnitz rule
contains the dynamical term and the Noether term, which we are used for the derivation of the conservation laws.

We remark here that from the point of view of mathematical definition $\delta{\mathcal L}/\delta{\eta}$ is a linear functional, i.e., in order to get a numerical value, one has to apply it  to the test function (we named it $\chi$ above), which does not have a small norm a priori.
%
\subsubsection{\label{ssubec:constr_var}Constrained Eulerian variations}
%
Before proceeding with the computation of the first variation of the Eulerian action functional (\ref{Euler_2_action}), we recall how the Vlasov field variations $\delta \widehat{F}$ are calculated.
In the passage from the Lagrangian to the Eulerian description, there exists a particle relabeling symmetry (meaning that we remove the label associated with each particle in the Lagrangian description of the plasma). The unconstrained variations on the relabeling transformation $\bm \psi({\bf Z}_0)={\bf Z}({\bf Z}_0,t)$ (where ${\bf Z}$ are the coordinates in phase space and ${\bf Z}_0$ the initial position of the particles in the Lagrangian description) translate in the constrained variations of the Vlasov distribution in the following form~\cite{Holm_Marsden_Ratiu_1998,Squire_Qin_2013}
$$
\delta \widehat{\mathcal F}\equiv \{\widehat{\cal S}, F\}_{\rm ext},
$$  
where $\widehat{\cal S}$ is an arbitrary function. 

For convenience, we separate the action functional into its field part (field) and Vlasov (Vl) parts :
\begin{eqnarray}
&&{\mathcal A}^{\mathcal E}[\phi_1,{\mathbf A}_1,{\mathcal F}]\equiv {\mathcal A}^{{\mathcal E},{(\mathsf{field})}}
[\phi_1,{\mathbf A}_1]+{\mathcal A}^{{\mathcal E},{(\mathsf{Vl})}}[\phi_1,{\mathbf A}_1,{\mathcal F}]
\label{A_E_separation}
\\
&&\qquad ={\mathcal A}^{{\mathcal E},{(\mathsf{field})}}_{\mathsf{el}}[\phi_1,{\mathbf A}_1]+
{\mathcal A}^{{\mathcal E},{(\mathsf{field})}}_{\mathsf{magn}}[\phi_1,{\mathbf A}_1]+
{\mathcal A}^{{\mathcal E},{(\mathsf{Vl})}}_{\mathsf{lin}}[\phi_1,{\mathbf A}_1,{\mathcal F}]+
{\mathcal A}^{{\mathcal E},{(\mathsf{Vl})}}_{\mathsf{nonlin}}[\phi_1,{\mathbf A}_1].
\nonumber
\end{eqnarray}
The field part is further divided into the electric part:
\begin{eqnarray}
{\mathcal A}^{{\mathcal E}, ({\mathsf{field}})}_{\mathsf{el}}\left[\phi_1,\mathbf{A}_{1}\right] \equiv
\frac{\epsilon_{\delta}^2}{8\pi}\int dV\  \left|\mathbf E_1\right|^2,
\end{eqnarray}
and the magnetic part:
\begin{eqnarray}
{\mathcal A}^{{\mathcal E}, ({\mathsf{field}})}_{\mathsf{magn}}\left[\phi_1,\mathbf{A}_{1}\right]\equiv
-\frac{1}{8\pi}\int dV\ \left|\mathbf{B}_0+\epsilon_{\delta}\mathbf{B}_1\right|^2.
\end{eqnarray}
Furthermore we separate the linear and nonlinear contributions to the Vlasov part in the following way:
\begin{eqnarray}
{\mathcal A}^{{\mathcal E}, ({\mathsf{Vl}})}_{\mathsf{lin}}\left[\phi_1,{\mathbf A}_{1},{\mathcal F}\right]\equiv 
-\int\ \ d^6 Z dw \ \mathcal{F}\ {\mathcal H}_{1},
\end{eqnarray}
and
\begin{eqnarray}
{\mathcal A}^{{\mathcal E}, ({\mathsf{Vl}})}_{\mathsf{nonlin}}\left[\phi_1,{\mathbf A}_{1}\right]\equiv 
-\epsilon_\delta^2 \int\ d^6 Z \ F_0\ H_2. 
\end{eqnarray}

%
\subsubsection{\label{ssubsec:fields_contrib}Fields contributions}
%

We start by calculating the functional derivatives of the field contributions to the action functional.  
\begin{equation*}
\delta{\mathcal A}^{{\mathcal E},({\mathsf{field}})}\equiv \frac{\delta\mathcal A^{{\mathcal E},({\mathsf{field}})}}{\delta\phi_1}\circ\widehat{\phi}_1
+\frac{\delta\mathcal A^{{\mathcal E},({\mathsf{field}})}}{\delta\mathbf A_1}\circ\widehat{\mathbf A}_1.
\end{equation*}
First, for the electrostatic field term with ${\mathbf E}_1$ we have:
\begin{eqnarray*}
\frac{\delta{\mathcal A}^{{\mathcal E}, ({\mathsf{field}})}_{\mathsf{el}}}{\delta\phi_1}\circ\widehat{\phi}_1&=&
\frac{d}{d\nu}\left.\left[\int\frac{dV}{8\pi}\ \epsilon_{\delta}^2\left|{\mathbf E}_1-\nu{\bm\nabla}\widehat{\phi}_1\right|^2\right]\right|_{\nu=0}
\nonumber
\\
&=&
\underbrace{-\epsilon_{\delta}^2\int \frac{dV}{4\pi} \ {\mathbf E}_1\cdot\bm\nabla\widehat{\phi}_1}_{\mathsf{Weak \ dynamics}}
\nonumber
\\
&=&
\underbrace{\epsilon_{\delta}^2\int \frac{dV}{4\pi}{\bm\nabla}\cdot{\mathbf E}_1\ \widehat{\phi}_1}_{\mathsf{Dynamical \ term}}-
\underbrace{\epsilon_{\delta}^2\int \frac{dV}{4\pi}{\bm\nabla}\cdot\left({\mathbf E}_1\ \widehat{\phi}_1\right)}_{\mathsf{Noether's \ term}},
\end{eqnarray*}
and
\begin{eqnarray*}
\frac{\delta{\mathcal A}^{{\mathcal E}, ({\mathsf{field}})}_{\mathsf{el}}}{\delta\mathbf{A}_{1}}\circ\widehat{\mathbf A}_1=
\frac{d}{d\nu}\left.\left[\int\frac{dV}{8\pi}\ \epsilon_{\delta}^2\left|{\mathbf E}_1-\nu\ \frac{1}{c}\partial_t\widehat{\mathbf A}_1\right|^2\right]\right|_{\nu=0}
\nonumber
=
\underbrace{-\epsilon_{\delta}^2\int \frac{dV}{4\pi} \ {\mathbf E}_1\cdot \frac{1}{c}\partial_t\widehat{\mathbf A}_1}_{\mathsf{Weak  \ dynamics}}
\\
\nonumber
=
\epsilon_{\delta}^2\int \frac{dV}{4\pi} \underbrace{\frac{1}{c}\ \partial_t{\mathbf E}_1\cdot\widehat{\mathbf A}_1}_{\mathsf{Dynamical\ term}}
-\epsilon_{\delta}^2\int \frac{dV}{4\pi} \underbrace{\ \frac{1}{c}\ \partial_t\left( {\mathbf E}_1\cdot \widehat{\mathbf A}_1\right)}_{\mathsf{Noether's\ term}}.
\end{eqnarray*}
Next, we derive a magnetic contribution with ${\mathbf B}_1=\bm\nabla\times{\mathbf A}_1$:
\begin{eqnarray}
\nonumber
\frac{\delta{\mathcal A}^{{\mathcal E}, ({\mathsf{field}})}_{\mathsf{mag}}}{\delta{\mathbf A}_1}\circ\widehat{\mathbf A}_1&=&
-\frac{d}{d\nu}\left.
\left[\int\frac{dV}{8\pi}\left|{\mathbf B}_0+\epsilon_{\delta}\bm\nabla\times\left({\mathbf A}_{1}+\nu\ \widehat{\mathbf A}_{1}\right)\right|^2\right]
\right|_{\nu=0}
\\
\label{delta_field_magnetic_A_1}
&=&
\underbrace{-\epsilon_{\delta}
\int \frac{dV}{4\pi}
\ \left({\mathbf B}_0+\epsilon_{\delta}{\mathbf B}_1\right)\cdot\left(\bm\nabla\times\widehat{\mathbf A}_1\right)}_
{\mathsf{Weak \ dynamics}}
\\
\nonumber
&=&
-\epsilon_{\delta}\int\frac{dV}{4\pi}\left(\underbrace{\widehat{\mathbf A}_1\cdot\bm\nabla\times\left({\mathbf B}_0+\epsilon_{\delta}{\mathbf B}_1\right)}_{\mathsf{Dynamical\ term}}
-\underbrace{\bm\nabla\cdot\left[\left({\mathbf B}_0+\epsilon_{\delta}{\mathbf B}_1\right)\times\widehat{\mathbf A}_1\right]}_{\mathsf{Noether's\ term}}\right).
\end{eqnarray}



%
\subsubsection{\label{ssubsec:vlasov_contrib}Vlasov contributions}
%
We proceed with the contributions from the Vlasov parts of the action functional. The first variation of the linear Vlasov part writes:
\begin{eqnarray}
\delta{\mathcal A}^{{\mathcal E}, ({\mathsf{Vl}})}_{\mathsf{lin}}&=&
\frac{\delta{\mathcal A}^{{\mathcal E}, ({\mathsf{Vl}})}_{\mathsf{lin}}}{\delta{\mathcal F}}\circ \delta\widehat{\mathcal F} +
\frac{\delta{\mathcal A}^{{\mathcal E}, ({\mathsf{Vl}})}_{\mathsf{lin}}}{\delta{\phi_1}}\circ \widehat{\phi}_1+
\frac{\delta{\mathcal A}^{{\mathcal E}, ({\mathsf{Vl}})}_{\mathsf{lin}}}{\delta{\mathbf A_{1}}}\circ \widehat{\mathbf A}_{1}
\label{Vlasov_contr}
\\
&=& 
-\int d^6 Z\ dw \left(\ {\mathcal H}_1\ \delta{\widehat{\mathcal F}}+{\mathcal F}\ \left(\frac{\delta{\mathcal H}_1}{\delta \phi_1}\circ\widehat{\phi}_1 +
\frac{\delta{\mathcal H}_1}{\delta \mathbf A_1}\circ\widehat{\mathbf A}_1\right)\ \right). 
\label{vlasov_variation}
\end{eqnarray}
To explicit the first contribution we are using the expression for the constrained Eulerian variation 
$\delta\widehat{\mathcal {F}}$. This expression can be further rewritten by using the Leibnitz rule for the Poisson bracket on the extended phase space such that the dynamical and the Noether contributions become:
\begin{eqnarray}
\frac{\delta{\mathcal A}^{{\mathcal E}, ({\mathsf{Vl}})}_{\mathsf{lin}}}{\delta{\mathcal F}}\circ \delta\widehat{\mathcal F}
&=&
-\int d^6 Z\ dw\ {\mathcal H}_1\ \{\widehat{\cal S},{\mathcal F}\}_{\mathsf{ext}}\label{Vlasov_var_1}
\\
&=&
-\int d^6 Z\ dw\ \underbrace{\{\widehat{\cal S} {\mathcal H}_1,{\mathcal F}\}_{\mathsf{ext}}}_{\mathsf{Noether's\ term}}+
\int d^6 Z\ dw\ \widehat{\cal S} \underbrace{\{{\mathcal H}_1,{\mathcal F}\}_{\mathsf{ext}}}_{\mathsf{Dynamical \ Vlasov}}.
\nonumber
\end{eqnarray}

The two remaining terms in Eq.~(\ref{vlasov_variation}) contribute to the gyrokinetic Maxwell's equations via polarization and magnetization terms.

Next we consider the contributions in the action functional from the nonlinear Vlasov part.
Since the second order reduced dynamics in the functional (\ref{Euler_2_action}) is associated  with the non-dynamical part of the distribution function $F_0$ only, it naturally leads to contributions in the dynamical equations and does not provide any Noether terms by construction:

\begin{eqnarray*}
\delta{\mathcal A}^{{\mathcal E}, ({\mathsf{Vl}})}_{\mathsf{nonlin}}
=
\frac{\delta{\mathcal A}^{{\mathcal E}, ({\mathsf{Vl}})}_{\mathsf{nonlin}}}{{\delta\phi_1}}\circ \widehat{\phi}_1+
  \frac{\delta{\mathcal A}^{{\mathcal E}, ({\mathsf{Vl}})}_{\mathsf{nonlin}}}{{\delta\mathbf A_{1}}}\circ \widehat{\mathbf A}_{1}
=
-\epsilon_\delta^2\int d^6 Z \ F_0\ \left(\frac{\delta{H}_2}{\delta\phi_1}\circ\widehat{\phi}_1 +
\frac{\delta{H}_2}{\delta\mathbf A_1}\circ\widehat{\mathbf A}_1\right).
\end{eqnarray*}

Now we show how the polarization and magnetization effects arise in the reduced Maxwell-Vlasov equations:
We notice that the reduced particle dynamics, contained in the Hamiltonians $H_1$ and $H_2$, is evaluated on the reduced phase space at the gyrocenter position $\mathbf{X}$, while the electromagnetic potentials $\phi_1$ and $\mathbf{A}_1$ are evaluated at the initial non-reduced space position $\mathbf r$. Therefore, the difference between both positions has to be systematically taken into account while evaluating the functional derivatives. This leads to the appearance of polarization and magnetization terms in the right hand side of the gyrokinetic Poisson and Amp\`ere equations.

As mentioned in the previous section, the first variation of the action functional can be rewritten in a form which contains two types of terms: those multiplied by the test functions $(\widehat{\phi}_1, \widehat{\mathbf A}_1,\widehat{\cal S})$, and other terms, representing exact derivatives with respect to time and space variables.
The first category of terms leads to the equations of motion, while the second one are used later for the derivation of conservation laws via Noether's theorem.

We provide an explicit derivation of the polarization contributions from the reduced particle dynamics in Sec.~\ref{sec:variational_explicit}. Here we proceed with the derivation of the reduced Maxwell-Vlasov equations and a brief presentation of the Noether procedure for the derivation of the energy conservation law.

%
\subsection{\label{subsec: implicit_eqns}Equations of motion: implicit weak form}
%

We write down the reduced Maxwell-Vlasov system corresponding to the Eulerian action functional (\ref{Euler_2_action}) in an implicit form, i.e., without specifying the expressions for the functional derivatives of the reduced Hamiltonians $H_1$ and $H_2$, essentially representing the polarization effects due to the dynamical reduction on the particle phase space. The explicit derivation is done in Sec.~\ref{sec:variational_explicit}.

We start by writing the implicit equations of motion in a \textit{weak} form (i.e., applied on the test functions), which is essential for the numerical implementation as well as for the derivation of energy conservation.
In order to write the \textit{weak} form of the second order gyrokinetic Maxwell - Vlasov equations we collect all the contributions to the first variation 
of the action functional.

Concerning the gyrokinetic Poisson equation, we have

\begin{eqnarray}
&&0=\frac{\delta{\mathcal A}^{\mathcal E}}{\delta\phi_1}\circ\widehat{\phi}_1\Rightarrow
\label{Poisson_implicit_weak}
\\
&&-\epsilon_{\delta}\int dV\ \left({\mathbf E}_1\cdot\bm\nabla\widehat{\phi}_1\right)=
4\pi\int d^6Z\ \left(F_0+\epsilon_{\delta}F_1\right)\frac{\delta H_1}{\delta\phi_1}\circ\widehat{\phi}_1+
4\pi\epsilon_\delta\int d^6Z F_0\ \frac{\delta H_2}{\delta\phi_1}\circ\widehat{\phi}_1,
\nonumber
\end{eqnarray}
and for the Amp\`ere equation, we have:
\begin{eqnarray}
\nonumber
&&0=\frac{\delta{\mathcal A}^{\mathcal E}}{\delta {\mathbf A}_1}\circ\widehat{\mathbf A}_1\Rightarrow
\\
\nonumber
&&\int dV\ 
\left[ 
\left({\mathbf B}_0+\epsilon_{\delta}{\mathbf B}_1\right)\cdot\bm\nabla\times\widehat{\mathbf A}_1
+\frac{\epsilon_{\delta}}{c}{\mathbf E}_1\cdot\partial_t\widehat{\mathbf A}_1
\right]=
\nonumber
\\
&&
-4\pi\int \ d^6Z \left(F_0+\epsilon_{\delta}F_1\right)\frac{\delta H_1}{\delta{\mathbf A}_1}\circ\widehat{\mathbf A}_1-
4\pi\epsilon_\delta\int \ d^6Z F_0\ \frac{\delta H_2}{\delta\mathbf A_1}\circ\widehat{\mathbf A }_1.
\label{Ampere_implicit_weak}
\end{eqnarray}

These equations can be rewritten in a strong form (i.e., without making apparent the test function explicitly). It can be achieved by integration by parts and then using the arbitrariness of the test functions.
We give an explicit expression for the strong form of these equations in the next section, where we make use of the explicit expressions for $H_1$ and $H_2$. 

Finally, using the arbitrariness of $\widehat{\cal S}$, the Vlasov equation is obtained from the extended phase space bracket using Eq.~(\ref{Vlasov_var_1}):
\begin{equation}
0=\frac{\delta{\mathcal A}^{{\mathcal E}, ({\mathsf{Vl}})}_{\mathsf{lin}}}{\delta{\mathcal F}}\circ \delta\widehat{\mathcal F}
\Rightarrow \{{\mathcal H}_1,{\mathcal F}\}_{\mathsf{ext}}=0.
\label{Vlasov_implicit}
\end{equation}
Equations~(\ref{Poisson_implicit_weak}), (\ref{Ampere_implicit_weak}) and (\ref{Vlasov_implicit}) define the dynamical gyrokinetic Maxwell-Vlasov equations in an implicit way (i.e., for generic $H_1$ and $H_2$).  

%
\subsection{{\label{subsec:Noether_method}}Noether method and energy conservation law}
%

Noether's method in classical field theory is used to associate the symmetries of the action functional with conserved quantities.
The general Noether's transport equation has the following form:
$$
\frac{\partial\mathsf S}{\partial t}+\bm{\nabla}\cdot{\mathbf J}=\delta {\mathcal L}^{\mathcal E}.
$$
where $\mathsf{S}$ is the Noether's density, ${\mathbf J}$ is the Noether's current and  ${\mathcal L}^{\mathcal E}$ is the Lagrangian density defined as
$$
{\mathcal A}^{\mathcal E}\equiv \int dV\ {\mathcal L}^{\mathcal E}.
$$
The variation $\delta{\mathcal L}^{\mathcal E}$ is defined accordingly to the conservation law we are deriving.

By collecting the Noether terms, which we have derived in Sec.~\ref{sec:IIIA}, we obtain the expressions for ${\mathsf S}$ and $\mathbf{J}$ suitable for conservation laws derivation:
\begin{eqnarray}
{\mathsf S}&=&-\frac{\epsilon_{\delta}^2}{4\pi}\frac{1}{c} {\mathbf E}_1\cdot\widehat{\mathbf A}_1 +
\int dW \ d w\ {\mathcal F}\ \widehat{\cal S},\label{Ndensity}
\\
{\mathbf J}&=&-\frac{\epsilon_{\delta}^2}{4\pi}\ {{\mathbf E}_1}\ \widehat{\phi}_1
\nonumber
\\
&+&
\frac{\epsilon_{\delta}}{4\pi}\left[\left({\mathbf B}_0+\epsilon_{\delta}{\mathbf B}_1\right)\times\widehat{\mathbf A}_1\right]+
\int d W\ d w\ 
{\mathcal F}\{{\mathbf X},{\cal H}_1\}_{\rm ext} \widehat{\cal S}, \label{Ncurrent}
\end{eqnarray}
where $dW= B_{\|}^*dp_z\ d\mu\ d\theta$.

The last term in the expression for Noether density $\mathsf S$ and Noether current $\mathbf{J}$ is obtained from the Vlasov part of the action functional, its explicit derivation from Eq.~(\ref{Vlasov_var_1}) are summarized in Appendix~\ref{app:Noether_method}.

Below we consider the derivation of the energy conservation.
The energy conservation is derived from performing infinitesimal time translations $t\rightarrow t+\delta t$ on the Eulerian action 
$\mathcal {A}^{\mathcal E}$. The explicit expression for the corresponding generating function $\widehat{\mathcal S}$, which also defines the constrained variations of the Vlasov field, is given by $\widehat{\cal S}=-w\ \delta t$.
The expressions of  electromagnetic field and Lagrangian density variations are defined as:
\begin{eqnarray}
\widehat{\phi}_1&=&-\delta t\ \frac{\partial\phi_1}{\partial t},
\label{time_var_phi}\\
\widehat{\mathbf{A}}_{1}&=&-\delta t\ \frac{\partial {\mathbf A}_{1}}{\partial t}=c \ \delta t\left( {\mathbf E}_{1}+\bm{\nabla}\phi_1\right),
\label{time_var_A}\\
\delta{\mathcal L}^{\mathcal E}&=&-\delta t \frac{\partial {\mathcal L}^{\mathcal E}}{\partial t}.
\label{time_var_L}
\end{eqnarray}

In the following section, we derive the explicit expression for the energy density, which contains FLR terms up to second order.
Since ${\cal H}=H-w$ and ${\cal F}=F\delta(w-H)$, the integral $\int dw {\cal F}{\cal H}$ vanishes. 
Therefore the time derivative of the Lagrangian density is given by:
\begin{eqnarray*}
\delta\mathcal L^{\mathcal E}=-\delta t\ \frac{\partial}{\partial t}
\left[
\frac{1}{8\pi}\left(\epsilon_{\delta}^2\right|{\mathbf E}_1\left|^2-
\right|{\mathbf B}_0+\epsilon_{\delta}{\mathbf B}_1\left|^2\right)-\int dW \ F_0 \ H_2
\right].
\end{eqnarray*}
It means that in the case of the energy conservation derived by infinitesimal time translations, the Maxwell part of the action as well as the truncated non-dynamical Vlasov part, contribute to the energy density $\mathsf{S}_E$. Its implicit expression is obtained from Eq.~(\ref{Ncurrent}):
\begin{eqnarray}
\mathsf{S}_E&=&\frac{1}{8\pi}\left(\epsilon_{\delta}^2\left|{\mathbf E}_1\right|^2+\left|{\mathbf B}_0+\epsilon_{\delta}{\mathbf B}_1\right|^2\right)+ \ \int dW \left((F_0+\epsilon_\delta F_1)(H_0+\epsilon_\delta H_1)+\epsilon_{\delta}^2 F_0\ H_2\right)\nonumber\\
&+& \frac{\epsilon_{\delta}^2}{4\pi}\ {\mathbf E}_1\cdot\bm\nabla\phi_1,
\label{Poisson_weak_Noether}\\
\mathbf{J}&=&
-\frac{\epsilon_{\delta}c}{4\pi}\left({\mathbf B}_0+\epsilon_{\delta}{\mathbf B}_1\right)\times{\mathbf E}_1
+\int dW\ (F_0+\epsilon_\delta F_1)(H_0+\epsilon_\delta H_1)\dot{\mathbf X}\nonumber
\\
&-&\frac{\epsilon_{\delta}^2}{4\pi}\left[\ {\mathbf E}_1\ \frac{\partial \phi_1}{\partial t}\ 
+c\left({\mathbf B}_0+\epsilon_{\delta}{\mathbf B}_1\right)\times\bm\nabla\phi_1\ \right].
\label{Ampere_weak_Noether}
\end{eqnarray}
Concerning the total energy defined as
$$
{\cal E}=\int dV\ \mathsf{S}_E,
$$
the integrals of the term (\ref{Poisson_weak_Noether}) can be rewritten using the weak form of the equation of motion (\ref{Poisson_implicit_weak}) with a test function $\widehat{\phi}_1\equiv\phi_1$. We introduce polarization and magnetization effects into the expression for the energy $\mathcal E$:
\begin{eqnarray*}
\mathcal{E}&=&\int dV\ \frac{1}{8\pi}\left(\epsilon_{\delta}^2\left|{\mathbf E}_1\right|^2+\left|{\mathbf B}_0+\epsilon_{\delta}{\mathbf B}_1\right|^2\right)+ \ \int d^6Z \left((F_0+\epsilon_\delta F_1)(H_0+\epsilon_\delta H_1) +\epsilon_{\delta}^2F_0\ H_2\right)\\
&-&\int d^6Z\ \epsilon_{\delta} \left[ \left(F_0+\epsilon_{\delta}F_1\right)\frac{\delta H_1}{\delta\phi_1}\circ\phi_1 + \epsilon_\delta F_0 \frac{\delta H_2}{\delta\phi_1}\circ\phi_1\right].
\end{eqnarray*}

The above expressions are valid for any $H_1$ and $H_2$. 
In the following section we use the expression of $H_1$ and $H_2$ obtained in Eqs.~(\ref{H1}) and (\ref{H2_full_GK}), in order to obtain explicit expressions for the equations of motion and the energy conservation laws in the case of the second order gyrokinetic reduction.

%
\section{\label{sec:variational_explicit} Eulerian second order action functional: explicit derivation}
%

The aim of this section is to provide explicit expressions for the reduced Maxwell-Vlasov equations in their \textit{weak} form [see Eqs.~(\ref{Poisson_implicit_weak})-(\ref{Ampere_implicit_weak})] and in their \textit{strong} form (i.e., with performing integration by parts and separating Noether's contributions) with in particular, specifying the two parts of the Hamiltonian $H_1$ and $H_2$, defined in the Sec.~\ref{subsec:gyrocenter}.

%
\subsection{\label{subsec:dynamic_noether}Dynamical and Noether terms}
%

By taking into account the assumptions on the electromagnetic fields and on the truncated particle dynamics discussed in Sec.~\ref{subsec:gyrocenter}, we substitute the expressions for $H_1$ and $H_2$ defined in Eqs.(\ref{H1})-(\ref{H2_full_GK}) into Eq.~(\ref{Euler_2_action}). We obtain an explicit expression for the Eulerian action:
\begin{eqnarray}
&&{\mathcal I_{\|}}^{\mathcal E}\left[\phi_1,A_{1\|},{\mathcal F}\right]\equiv \int_{t_1}^{t_2}{\mathcal A_{\|}}^{\mathcal E}\left[\phi_1,A_{1\|},{\mathcal F}\right]\ dt
\label{Eul_explicit}
\\
&=&
\int\frac{dV\ dt}{8\pi}\left(\epsilon_{\delta}^2\left|\bm\nabla_{\perp}\phi_1\right|^2-\left|{\mathbf B}_0+\epsilon_{\delta}\ \bm\nabla\times\left(A_{1\|}\widehat{\mathbf b}\right)
\right|^2\right)
\label{Maxwell}
\\
&-&\int d^8 Z \ \mathcal{F}
\left[H_0+\epsilon_{\delta}\ e\left\langle\phi_{\mathrm{1 gc}}\right\rangle-\epsilon_{\delta}\ e\ \frac{p_{z}}{m c}
\left\langle
A_{1 \| \mathrm{gc}}\right\rangle-w\right]
\label{Polarization_Vlasov}
\\&+&
\frac{\epsilon_{\delta}^2}{2}\ \int d^6 Z\ dt\  \frac{m c^2}{B^2}\ F_0\ \left|\bm\nabla_{\perp}\phi_1-\frac{p_{z}}{m c}\bm\nabla_{\perp}A_{1\|}\right|^2
\label{Polarization_1}
\\
&-&
\frac{\epsilon_{\delta}^2}{2}\ \int d^6 Z\ dt\  F_0\ \left(\frac{1}{m}\left(\frac{e}{c}\right)^2 A_{1\|}^2+
\frac{\mu}{B}\left|\bm\nabla_{\perp}A_{1\|}\right|^2+
{\color{black}\frac{\mu}{B}A_{1\|}\bm\nabla_{\perp}^2A_{1\|}}\right),
\label{Polarization_magnetic}
\end{eqnarray}
where the extended Vlasov function is given by Eq.~(\ref{F_tr_ext}).

We are following the same procedure for the first variation calculation as in Sec.~\ref{sec:IIIA}, but this time, with the explicit expressions for $H_1$ and $H_2$:

\begin{eqnarray}
\delta{\mathcal I_{\|}}^{\mathcal E}\equiv \int_{t_1}^{t_2} \delta\mathcal{A}^{\mathcal E} \left[\phi_1, A_{1\|},{\mathcal F}\right] dt
=
\int_{t_1}^{t_2}\left(
\frac{\delta{\mathcal A_{\|}}^{\mathcal E}}{\delta\phi_1}\circ\widehat{\phi}_1+
\frac{\delta{\mathcal A_{\|}}^{\mathcal E}}{\delta {A}_{1\|}}\circ\widehat{A}_{1\|}+
\frac{\delta{\mathcal A_{\|}}^{\mathcal E}}{\delta {\mathcal F}}\circ\delta\widehat{\mathcal F}
\right) dt.
\label{first_var_implicit}
\end{eqnarray}
We provide the details of the calculations in Appendix~\ref{app:Eulerian_explicit}.
Here we focus on the final form for the gyrokinetic Maxwell-Vlasov equations in \textit{weak} and \textit{strong} forms as well as the expression for the conserved energy density.

%
\subsection{\label{subsec:Eulerian_eqns}Equations of motion}
%
%
\subsubsection{\label{ssubsec:gk_vlasov}Gyrokinetic Vlasov equation}
%
The Vlasov equation follows from the variational principle in the form of an exact derivative accordingly to Eq.~(\ref{Vlasov_implicit}).
This is equivalent to the statement that the Vlasov equation is reconstructed from the first order gyrocenter characteristics (\ref{GY_CHAR_1}). 
This equation explicitly writes:
\begin{eqnarray}
\frac{\partial F_1}{\partial t}&=&-\{F_1,H_0\}_{\mathrm{gc}}-\{F_0,H_1\}_{\mathrm{gc}}-\epsilon_{\delta}\{F_1,H_1\}_{\mathrm{gc}},
\label{Vlasov_Eulerian}
\end{eqnarray}
where the two first terms represent the linear drive in the system (first term: coupling between the background dynamics and the dynamical part of the Vlasov field; second term: coupling between the background (non-dynamical) distribution with the first order fluctuations), the last term represents the non-linear coupling between the dynamical part of the Vlasov field with the first order Hamiltonian.

%
\subsubsection{\label{ssubsec:gk_Poisson}Gyrokinetic Poisson equation}
%

From Eq.~(\ref{Poisson_implicit_weak}), using the explicit expressions for the functional derivatives of $H_1$ and $H_2$ given by Eqs.~(\ref{delta_H_1_delta_phi_1}) and (\ref{delta_H_2_delta_phi_1}), the Poisson equation in \textit{weak} form writes:
\begin{eqnarray*}
\frac{\epsilon_{\delta} }{4\pi}\int dV\ \bm\nabla_{\perp} \phi_1\cdot\bm\nabla_{\perp} \widehat{\phi}_1
&=&
-\epsilon_{\delta}\int dV\ dW\ \ \left(\frac{m c^2}{B^2}\ F_0\right) \left[\bm\nabla_{\perp}\phi_1-\frac{p_{z}}{m c}\bm\nabla_{\perp}A_{1\|}\right]\cdot \bm\nabla_{\perp}\widehat{\phi}_1
\\
&+&e\
\int\ dV\ dW \ \left(F_0+\epsilon_{\delta}F_1\right)\
\left\langle\delta^3(\mathbf{X}+\bm\rho_0-{\mathbf r})\ \widehat{\phi}_1\right\rangle.
\end{eqnarray*}
Here we define the guiding-center gyro-averaging operator $\mathcal J_0^{\mathrm{gc}}$ as:
\begin{eqnarray}
\left\langle\delta^3(\mathbf{X}+\bm\rho_0-{\mathbf r})\ \widehat{\phi}_1\right\rangle
&\equiv&
\frac{1}{2\pi}\int_0^{2\pi} d\theta \ \delta^3(\mathbf{X}+\bm\rho_0-{\mathbf r})\ \widehat{\phi}_1({\mathbf r}),
\\
&=&
\frac{1}{2\pi}\int_0^{2\pi}\  d\theta\  \widehat{\phi}_1\left(\mathbf{X}+\bm\rho_0\right)\equiv {\mathcal J}_0^{\mathrm{gc}} \left(\widehat{\phi}_1\right),
\label{J0_gc}
\end{eqnarray}
and as a consequence, 
\begin{eqnarray*}
\int dV\ dW \ \left(F_0+\epsilon_{\delta}F_1\right)\
 \left\langle\delta^3(\mathbf{X}+\bm\rho_0-{\mathbf r}) \ \widehat{\phi}_1\right\rangle\equiv
 \int dV\ dW \ \left(F_0+\epsilon_{\delta}F_1\right)\
 \mathcal{J}_0^{\mathrm{gc}}\left(\widehat{\phi}_1\right).
 \label{J_0}
\end{eqnarray*}
Integrating by parts, (see Appendix~\ref{app:Eulerian_explicit}, Eq.~(\ref{dyn_polmix}) for details) and since the weak form is valid for any arbitrary function $\widehat{\phi}_1$, we have:
\begin{eqnarray}
0=  
 \frac{\epsilon_{\delta}}{4\pi}\bm\nabla_{\perp}^2\phi_1 
+
\epsilon_{\delta}\ \int \ dp_{z}\ d\mu \ \bm\nabla_{\perp}\cdot
\left[
\frac{mc^2}{B^2}B_{\|}^* \ F_0\ \bm\nabla_{\perp}
\left(
\phi_1-\frac{p_{z}}{m c}A_{1\|}
\right)
\right]
\nonumber
\\
+ e
\int \ dp_{z}\ d\mu\ 
B_{\|}^*\ \mathcal J_0^{{\mathrm{gc}}\dagger} \left(F_0+\epsilon_{\delta} F_1\right).
\label{Poisson_full}
\end{eqnarray}
This is the explicit \textit{strong} form of the Poisson equation. We notice that the requirement for the gyro-averaging operator being Hermitian, i.e.,
$\mathcal{J}_0^{\mathrm{gc}}=\mathcal{J}_0^{\dagger\mathrm{gc}}$ is not necessary in case of the finite-element discretization performed for the construction of a PIC code, because in that case we are discretizing equations in their \textit{weak} form and we do not need to shift the gyro-averaging operator from the test function $\widehat{\phi}_1$ to the distribution function. However, an example of an Hermitian gyro-averaging operator can be found in Ref.~\cite{McMillan2012}.
%
\subsubsection{\label{ssubsection:gk_Ampere}Gyrokinetic Amp\`ere equation}
%
From Eq.~(\ref{Ampere_implicit_weak}), and using the expressions for the functional derivatives of $H_1$ and $H_2$, given by (\ref{delta_H_1_delta_A_1}), (\ref{delta_H_2_delta_A_1_polmix}) and (\ref{delta_H_2_delta_A_1}), the \textit{weak} formulation of the Amp\`ere equation writes:
\begin{eqnarray}
0&=&-\int \frac{dV}{4\pi}\  \epsilon_{\delta}
(\mathbf{B}_0+\epsilon_{\delta}{\mathbf B}_1)\cdot\bm\nabla\times\left(\widehat{A}_{1\|}\widehat{\mathbf b}\right)
\\
\nonumber
&-&\epsilon_{\delta}^2\int dV \ dW\ \frac{m c^2}{B^2}F_0
(\bm\nabla_{\perp}\phi_1-\frac{p_z}{mc}\bm\nabla_{\perp}A_{1\|}
)\cdot\left(\frac{p_z}{m c}\bm\nabla_{\perp}\widehat{A}_{1\|}\right)
-\epsilon_{\delta}^2\int dV\ dW\ F_0\ \left(\frac{e^2}{mc^2}A_{1\|}\widehat{A}_{1\|}
\right.
\\
\nonumber
&+&\left.\frac{\mu}{B}
\left[
\bm\nabla_{\perp}A_{1\|}\cdot\bm\nabla_{\perp}\widehat{A}_{1\|}
+A_{1\|}\bm\nabla_{\perp}^2\widehat{A}_{1\|}+
\widehat{A}_{1\|}\bm\nabla_{\perp}^2A_{1\|}
\right]
\right)
+
\epsilon_{\delta}\int \ dV\ dW \ \left(F_0+\epsilon_{\delta}F_1\right)\ 
\frac{p_z}{m c}\mathcal J_0^{\mathrm{gc}}\left(\widehat{A}_{1\|}\right).
\end{eqnarray}
We notice that the choice of a parallel magnetic potential $A_{1\|}$ naturally leads to the derivation of only the parallel component of the \textit{strong} gyrokinetic Amp\`ere equation:
\begin{eqnarray}
\nonumber
&&
0=
-\frac{1}{4\pi} \widehat{\mathbf b}\cdot\bm\nabla\times\left({\mathbf B}_0+\epsilon_{\delta} {\mathbf B}_1\right)+
\epsilon_{\delta} \ \int\ d\mu\ dp_{z} \bm\nabla_{\perp}\cdot
\left[
\frac{m c}{B^2}\ B_{\|}^* \ F_0\ \frac{p_{z}}{m}
 \bm\nabla_{\perp}\left(
\phi_1-\frac{p_{z}}{m c}A_{1 \|}
\right)
\right]
\label{Ampere_full}
\\
&&
-\epsilon_{\delta}\ \frac{e^2}{m c^2}\int \ dW\ F_0\ A_{1 \|}+
\epsilon_{\delta}\int\ d\mu\ dp_{z}  \bm\nabla_{\perp}\cdot
\left(
\frac{\mu}{B}B_{\|}^* \ F_0 
 \bm\nabla_{\perp}A_{1\|}\right)
\\
&&
-\frac{\epsilon_{\delta}^2}{2}\int \ d\mu\ dp_{z}\ \bm\nabla^2_{\perp}\left(\frac{\mu}{B}\ B_{\|}^*F_0 A_{1\|}\right)
-\frac{\epsilon_{\delta}^2}{2}\int \ d\mu\ dp_{z}\ \left(\frac{\mu}{B}\ B_{\|}^*F_0\right) \bm\nabla^2_{\perp}A_{1\|}
\nonumber
\\
\nonumber
&+&\int dW\ \frac{e\ p_{z}}{m c}\ 
\mathcal J_0^{{\mathrm{gc}}\dagger} 
\left(
F_0+\epsilon_{\delta} F_1
\right),
\end{eqnarray}
where we have taken into account Eqs.~(\ref{dyn_polmix_magn}), (\ref{Vl_polmag1}) and (\ref{Vl_polmag2}).
%
\subsection{\label{subsec:energy_cons}Conservation law for the energy}
%

By substituting the variations associated with time translations [see Eqs.~(\ref{time_var_phi}), (\ref{time_var_A}) and (\ref{time_var_L})] in the general expressions for the Noether density ${\mathsf S}$, using the equations of motion associated with ${\mathcal A}^{\mathcal E}$, derived into the previous section, we get the expression for the energy density in its implicit form:
\begin{eqnarray}
{\mathsf S}_E&=&\frac{1}{8\pi}\left(\epsilon_{\delta}^2\left|{\mathbf E}_1\right|^2+\left|{\mathbf B}_0+\epsilon_{\delta}{\mathbf B}_1\right|^2\right) -\frac{\epsilon_\delta^2}{4\pi} |\bm\nabla_\perp \phi_1|^2\nonumber \\
&+&\int dW\ \left( (F_0+\epsilon_\delta F_1)(H_0+\epsilon_\delta H_1)+\epsilon_\delta^2 F_0 H_2\right),
\label{E_tr}
\end{eqnarray}
leading to the expression for the total energy
\begin{eqnarray}
{\mathcal E}_{\mathrm 2}=
\int dV\ dW \
\left(
F_0+\epsilon_{\delta} F_1
\right)
\left(
H_0-
\epsilon_{\delta}\ e\frac{p_{z}}{m}
\left\langle
A_{1\| gc}
\right\rangle
\right)
\\
+\nonumber
\frac{\epsilon_{\delta}^2}{2}\int dV\ dW \ 
F_0 \left(
\frac{e^2}{c^2}\frac{1}{m}A^2_{1\|}+\frac{\mu}{B}
\left\vert
\bm\nabla_{\perp}A_{1\|}
\right\vert^2+
\frac{\mu}{B}
A_{1\|}\bm\nabla_{\perp}^2 A_{1\|}
\right)
\\
+\nonumber
\frac{\epsilon_{\delta}^2}{2}\int dV\ dW \ F_0 \
\frac{m c^2}{B^2}
\left(
\left|
\bm\nabla_{\perp}\phi_1\right|^2-
\left(\frac{p_{z}}{m c}\right)^2\left|\bm\nabla_{\perp} A_{1\|}
\right|^2
\right)
\\
\nonumber
+\frac{1}{8\pi}\int dV
\left(
\epsilon_{\delta}^2 |\bm\nabla_{\perp}\phi_{1}|^2+
|{\mathbf B}_0+\epsilon_{\delta}{\mathbf B}_1
|^2
\right).
\end{eqnarray}

We remark, that in the electromagnetic case, there is a part of the energy provided by the background and fluctuating magnetic field. Therefore the field energy contribution can not be completely removed using the quasi-neutrality approximation, as it is possible in the electrostatic case. The field part of the energy should then be included into the code diagnostics.
%
\section{\label{sec:ORB5_variational} Eulerian variational principle for the ORB5 code model}
%
 In the previous sections, we have explicitly derived the equations of motion and the energy density corresponding to the Eulerian action functional 
(\ref{Eul_explicit}), which contains up to $\mathcal{O}(\epsilon_{\delta}^2)$ terms together with the second order FLR corrections. In this section we rewrite the second order Eulerian variational functional in a more compact form and then perform on it all necessary approximations in order to be able to derive the gyrokinetic Maxwell-Vlasov system of equations currently implemented in ORB5. We also aim to compare the energy density corresponding to the variational principle with the diagnostics of the code.
Our main goal here is to compare the gyrokinetic Maxwell-Vlasov models coming from a different first principle.

%
\subsection{Second order action functionals}
%

To get the ORB5 code model, several physical approximations are performed on the action functional (\ref{Eul_explicit}).  

%
\subsubsection{\label{ssubsec:quasi_neutral}The quasi-neutrality approximation}
%

We start by considering the most common physical assumption: the quasi-neutrality approximation, which is implemented in ORB5. 
The quasi-neutrality approximation allows one to neglect the $\left|\mathbf{E}_1\right|^2$ term in the Maxwell's part of the Eulerian action (\ref{Eul_explicit}).
This term is usually small compared to the second order polarization terms contained in $H_2$.
We recall that standard gyrokinetic ordering pushes the parallel component of electric field  at the next order compared to its perpendicular component $|E_{1\|}|\sim\epsilon_{\delta}|\mathbf{E}_{1\perp}|$.  The Eulerian action (\ref{Eul_explicit}) does not contain the parallel component of the electric field. In addition to this we are now taking into account the separation of the characteristic spatial scales, resulting from the fact that the ion sound Larmor radius $\rho_s$ is larger than the Debye length 
$\lambda_d$:
\begin{equation*}
\frac{\rho_s^2}{\lambda_d^2}=\frac{8\pi n m c^2}{B^2}=\frac{c^2}{v_A^2}\gg1,
\end{equation*}
where $v_A$ is the Alfv\`en velocity. In the electrostatic approximation, we have
\begin{eqnarray*}
\int \frac{d V}{8\pi}|\bm\nabla_{\perp}\phi_1|^2+\int dV\ dW \ F_0\ \frac{m c^2}{2 B^2}|\bm\nabla_{\perp}\phi_1|^2=
\frac{1}{8\pi}\int dV\ \left(1+\frac{\rho_s^2}{\lambda_d^2}\right)|\bm\nabla_{\perp}\phi_1|^2.
\end{eqnarray*}
Therefore, the electrostatic contribution to the Maxwell's part of the action functional is omitted.

%
\subsubsection{\label{ssubsec:magn_field_approx}Perturbed magnetic field approximation}
%
The next approximation concerns again the Maxwell's part of the action, and more precisely the perturbed part of the magnetic field.
Most of the physical models omit the curvature contributions of the perpendicular part of magnetic field perturbation $\mathbf B_1$, i.e., the term
${\mathbf B}_{\perp}=\widehat{\mathbf b}\times\bm\nabla A_{\|}$. This approximation leads to the violation of the constraint of divergence free magnetic field at the second order in $\epsilon_B$, referred earlier to as the small parameter related to the background field non-uniformities. Regardless of this, the term
$$
-\epsilon_\delta\int \frac{dV}{8\pi}{\bf B}_0\cdot \bm\nabla\times (A_{1\| \mathrm{gc}}\hat{\bf b}),
$$
is neglected in the action functional (\ref{Eul_explicit}). 

%
\subsubsection{\label{ssubsec:particle_approx}Particle dynamics approximation}
%

The last approximation refers to the particle part of the action functional.
To get the ORB5 model, which in its current form, does not take into account the coupling between the reduced Poisson and Amp\` ere equations, the second order Hamiltonian $H_2$ should not contain any "mixed" electromagnetic potential perturbation.
This is why the ORB5 model uses linear polarization approximation, and the $H_2$ Hamiltonian considered in the code is:
\begin{equation}
H_2^{\mathsf{ORB5}} \equiv \frac{e^2}{2 mc^2}\left\langle A_{1\| \mathrm{gc}}\right\rangle^2-\frac{mc^2}{2 B^2}\left|\bm\nabla_{\perp}\phi_1\right|^2,
\label{H2_nemo_GK}
\end{equation}
which is different from $H_2$ given by Eq.~(\ref{H2_full_GK}) obtained by direct derivation in the framework of the gyrokinetic reduction.
The final expression of the action functional leading to the gyrokinetic Maxwell-Vlasov equations corresponding to the ORB5 model is:
\begin{eqnarray}
\mathcal{I}^{\mathcal E}_{\mathrm{ORB5}}&\equiv&\int_{t_1}^{t_2} {\mathcal A}^{\mathcal E}_{\mathsf{\ ORB5}}\ dt
=
-\frac{\epsilon_{\delta}^2}{8\pi} \int dV dt\ \left|\bm\nabla_{\perp}A_{1\|}\right|^2
\nonumber\\
&-&\sum_{\mathrm{sp}}\int d^8 Z \ \mathcal{F}\ \mathcal{H}_1 - \sum_{\mathrm{sp}}\int d^6 Z dt \ F_0\ H_2^{\mathrm{ORB5}}.
\label{nemo_lagr_action}
\end{eqnarray}

%
\subsection{\label{subsec:ORB5_model}ORB5 Maxwell-Vlasov model}
%
In order to make a comparison between the gyrokinetic Maxwell-Vlasov equations obtained from the action functional~(\ref{Eul_explicit}) and those currently implemented in ORB5, we truncate the FLR expansion in Eq.~(\ref{H2_nemo_GK}) for the gyro-averaged magnetic potential $\left\langle A_{1\| \mathrm{gc}}\right\rangle$  at the second order.
We use the Taylor expansion in the vicinity of the reduced position $\mathbf{X}$ and we keep the guiding-center polarization corrections up to the second order in $\rho_0$ given by Eq.~(\ref{Second_FLR_A_1}).
Therefore, the first term in Eq.~(\ref{H2_nemo_GK}) is:
\begin{equation*}
\left\langle A_{1\| \mathrm{gc}}\right\rangle^2=\left(A_{1\|}+\frac{1}{2}\left\langle{\bm\rho}_0{\bm\rho}_0\right\rangle :\bm\nabla\bm\nabla A_{1\|}\right)^2=
A_{1\|}^2+m \left(\frac{c}{e}\right)^2\ \frac{\mu}{B} A_{1\|} \  \bm\nabla_{\perp}^2 A_{1\|}, 
\end{equation*}
whereas Eq.~(\ref{H2_full_GK}) contains an additional first order FLR contribution:
\begin{eqnarray}
\left\langle A_{1\| \mathrm{gc}}^2\right\rangle&=&\left\langle \left( A_{1\|}^2+{\bm\rho}_{0}\cdot\bm\nabla A_{1\|}+\frac{1}{2}{\bm\rho}_0{\bm\rho}_0:\bm\nabla\bm\nabla A_{1\|}\right)^2\right\rangle
\nonumber
\\
&=&A_{1\|}^2+m \left(\frac{c}{e}\right)^2\ \frac{\mu}{B}\left\vert\bm\nabla_{\perp}A_{1\|}\right\vert^2+m \left(\frac{c}{e}\right)^2\ \frac{\mu}{B}A_{1\|} \  \bm\nabla_{\perp}^2 A_{1\|}.
\end{eqnarray}

In addition, comparing Eqs.~(\ref{H2_full_GK}) and (\ref{H2_nemo_GK}), we notice that the first one contains electromagnetic corrections, while the second one is restricted to the electrostatic corrections only. The presence of electromagnetic contributions in $H_2$ results in a coupling between the reduced Poisson (quasi-neutrality) and Amp\`ere equations. The coupling of the gyrokinetic Maxwell equations can be of particular interest for further numerical studies with a Maxwellian initial distribution in the canonical variables (asymmetric background distribution in the physical variables). Such an implementation has a particular interest for the investigation of energetic particles.

%
\subsection{\label{subsec:ORB5_eqns_of_motion}Equations of motion}
%
%
\subsubsection{\label{ssubsec:ORB5_quasineutrality}Quasi-neutrality equations}
%
We start by comparing the quasi-neutrality equations. The first one is obtained from the second order Eulerian action~(\ref{Eul_explicit}) and in particular from Eq.~(\ref{Poisson_full})
\begin{eqnarray*}
&-&\epsilon_{\delta}\sum_{\mathrm{sp}}\int dW\frac{1}{B_{\|}^*}\ \bm\nabla_{\perp}\cdot \left[B_{\|}^*F_0\frac{m c^2}{B^2}\bm\nabla_{\perp}\left(\phi_1-
{\color{black}\frac{p_{z}}{m c}A_{1\|}}\right)\right]
= \sum_{\mathrm{sp}}e \int dW\ \mathcal J_0^{{\mathrm{gc}}\dagger} \left({\color{black}F_0}+\epsilon_{\delta} F_1\right),
\end{eqnarray*}
where the term in $\bm\nabla_\perp ^2 \phi_1$ has been neglected to reflect the same hypothesis formulated in ORB5. 
The quasi-neutrality equation used in ORB5 is:
\begin{eqnarray*}
-\sum_{\mathrm{sp}}\int dW\frac{1}{B_{\|}^*}\bm\nabla_{\perp}\cdot \left[B_{\|}^*F_0\frac{m c^2}{B^2} \bm\nabla_{\perp}\phi_1\right]
=\sum_{\mathrm{sp}}\ e \int dW\ \mathcal J_0^{{\mathrm{gc}}\dagger} F_1. 
\end{eqnarray*}

%
\subsubsection{\label{ssubsec:ORB5_Ampere}Amp\`ere's equations}
%
Taking into account the same approximation as in the ORB5 model with $\mathbf{B}_1=\widehat{\mathbf b}\times\bm{\nabla}A_{1\|}$, the Amp\`ere equation which follows from the Eulerian action~(\ref{Eul_explicit}) is obtained from Eq.~(\ref{Ampere_full}) and is given by:
\begin{eqnarray}
\frac{1}{4\pi} \bm\nabla^2_{\perp}A_{1\|}&=&
\int\ dW\ \frac{e p_{z}}{m c}\ \mathcal J_0^{{\mathrm{gc}}\dagger} F_1
-\int dW\ \frac{e^2}{m c^2}\left(A_{1\|} F_0\right)
+\int dW \frac{1}{B_{\|}^*}\bm\nabla_{\perp}\cdot \left[\left(F_0\frac{\mu}{B}B_{\|}^*\right)\bm\nabla_{\perp}A_{1\|}\right]
\nonumber
\\
\nonumber
&-&\frac{1}{2}\int dW\left(F_0\frac{\mu}{B}\right)\bm\nabla^2_{\perp} A_{1\|}
-\frac{1}{2}\int dW \frac{1}{B_{\|}^*}\bm\nabla^2_{\perp}\left(F_0\frac{\mu}{B}B_{\|}^* A_{\|}\right)
\\
&+&\int dW \frac{1}{B_{\|}^*} \bm\nabla_{\perp}\cdot\left[\left(F_0 B_{\|}^* \frac{c p_z}{B^2}\right)\bm\nabla_{\perp}\left(\phi_1-\frac{p_z}{mc} A_{1\|}\right)\right],
\label{Ampere_2_theory}
\end{eqnarray}
while the Amp\`ere equation which follows from the ORB5 action~(\ref{nemo_lagr_action}) is:
\begin{eqnarray}
\frac{1}{4\pi} \bm\nabla^2_{\perp}A_{1\|}&=&
\int dW\ \frac{e p_{z}}{m c}\ \mathcal J_0^{{\mathrm{gc}}\dagger} F_1
-\int dW\ \frac{e^2}{m c^2}\left(A_{1\|} F_0\right)
-\frac{1}{2}\int dW\left(F_0\frac{\mu}{B}\right)\bm\nabla^2_{\perp} A_{1\|}
\nonumber
\\
&-&\frac{1}{2}\int dW \frac{1}{B_{\|}^*}\bm\nabla^2_{\perp}\left(F_0\frac{\mu}{B}B_{\|}^* A_{\|}\right).
\label{Ampere_phys}
\end{eqnarray}
We notice that Amp\`ere's law obtained from Eulerian action~(\ref{Eul_explicit}) neglects the coupling with the electrostatic potential $\phi_1$. The third term in Eq.~(\ref{Ampere_2_theory}) is not present into Eq.~(\ref{Ampere_phys}), this is due to the differences identified in the expressions for the second order Hamiltonians $H_{2}$ and $H_2^{\mathrm{ORB5}}$. 
By neglecting the second order gradient of the background quantities, in particular the term proportional to $\bm\nabla^2_\perp\left(B_{\|}^*\frac{\mu}{B}F_0\right)$, combining the last two terms in Eq.~(\ref{Ampere_phys}), we get:
$$
-{\color{black}\int dW\ \bm\nabla_\perp\cdot\left(\frac{\mu}{B} F_0 \bm\nabla_{\perp} A_{1\|}\right)},
$$
which is more convenient for the numerical implementation in the PIC code.
The final Amp\`ere's law for ORB5 neglects the third term in the right hand side of Eq.~(\ref{Ampere_phys}) and it becomes:
\begin{equation*}
\sum_{\mathrm{sp}}\frac{1}{d_{\mathrm{sp}}^2} A_{1\|}+\sum_{\mathrm{sp}\neq{\mathrm{e}}}\ \nabla
\cdot\left(\beta_{sp}\bm\nabla_{\perp}A_{1\|}\right)+\bm\nabla_{\perp}^2 A_{1 \|}
=
\sum_{\mathrm{sp}} 4\pi\int dW\ \frac{e\ p_{z}}{m c} \mathcal J_0^{{\mathrm{gc}}\dagger}\ F_1,
\end{equation*}
where we have defined $d_{\mathrm{sp}}\equiv\frac{4\pi e^2 n_{0 e}}{m c^2}$ and $\beta_{\mathrm{sp}}=\frac{8\pi \mu n_{\mathrm{sp}}}{B^2}\equiv
\frac{8\pi n_{\mathrm{sp}} k_B T_{\mathrm{sp}}}{B^2}$.

\subsubsection{Vlasov equation}

The full second order Eulerian action provides the Vlasov equation with non-linear drive terms (\ref{Vlasov_Eulerian}), while the Eulerian action functional containing the physical reduction uses the first order gyrocenter characteristics (\ref{GY_CHAR_1}) to reconstruct the Vlasov equation and does not contain the nonlinear drive terms.

The following Vlasov equation is solved in ORB5:
$$
\frac{d {\cal F}}{dt}\equiv \left\{{\mathcal H}_1,{\mathcal F}\right\}_{\mathrm{ext}}=0,
$$
which represents basically the same equation as Eq.~(\ref{Vlasov_Eulerian})  where we have taken into account that  the background distribution is non-dynamical $\left\{F_0,H_0\right\}_{\mathrm{gc}}=0$ and static $\partial_t F_0=0$, which leads to the following:
\begin{eqnarray}
\frac{d F_1}{d t}=-\left\{F_0, H_1\right\}_{\mathrm{gc}}. 
\end{eqnarray}
In other words, the dynamics of the dynamical part of the distribution function $F_1$ is defined from the linear evolution of the background distribution $F_0$.
%
\section{\label{sec:Energy_conservation_ORB5}Energy conservation diagnostics}
%
The energy diagnostics implemented in the electromagnetic version of ORB5 are derived from the energy conservation law, which corresponds to the electromagnetic Lagrangian (\ref{nemo_lagr_action}).
Intuitively, the conserved energy density can be written as follows: 
\begin{eqnarray}
\mathcal{E}^{\mathsf{ORB5}}
&=&\sum_{\mathrm{sp}}\int dW\ dV\ H_0\ \left(F_0+\epsilon_{\delta}F_1\right)+ \sum_{\mathrm{sp}}\int dW\ dV\ H_1\ \left(F_0+\epsilon_{\delta}F_1\right) 
\nonumber
\\
&+&\sum_{\mathrm{sp}}\int dW\ dV\ H_2^{\mathsf{ORB5}}\ F_0+\int dV\ \frac{|\bm\nabla_{\perp} A_{1\|}|^2}{8\pi},
\label{ORB5_Energy}
\end{eqnarray}
which is equivalent to the result coming out from the direct Noether method in the framework of the Eulerian variational principle, once we have taken into account the Poisson equation corresponding to the truncated Lagrangian. Below we give a detailed explanation.

First, we explicitly write the expression for the second term
\begin{eqnarray*}
&&\int dV\ dW\ H_1 \ \left(F_0+\epsilon_{\delta}F_1\right) =
\int dV\ dW\  \left(F_0+\epsilon_{\delta}F_1\right) \mathcal J_0^{\mathrm{gc}}\left(\phi_1\right)
\\
&& \qquad -\int dV\ dW\ \frac{e p_z}{m c}\ \left(F_0+\epsilon_{\delta}F_1\right) \mathcal J_0^{\mathrm{gc}}\left(A_{1\|}\right),
\end{eqnarray*}
where we have used the explicit definition for the gyrocenter gyro-averaging operator (\ref{J0_gc}) to expand the expression for $H_1$ as follows:
$\left\langle\phi_{1 \mathrm{gc}}\right\rangle\equiv\left\langle\phi_1(\mathbf{X}+\bm\rho_0)\right\rangle=
\left\langle\phi_1(\mathbf r)\delta^3\left({\mathbf X+\bm\rho_0-\mathbf r}\right)\right\rangle\equiv\mathcal J_0^{\mathrm{gc}}\left(\phi_1\right)$
Next, using the quasi-neutrality equation and integrating by parts:
\begin{eqnarray}
\sum_{\mathrm{sp}} \int d V\ d W\ F_0\ \frac{m c^2}{B^2}\left|\bm\nabla_{\perp}\phi_1\right|^2  
&=&
-\sum_{\mathrm{sp}}\int d V \ d W \mathcal J^{{\mathrm{gc}}\dagger}_0 \left(F_0+\epsilon_{\delta} F_1\right)\ \phi_1,
\\
&=&
-\sum_{\mathrm{sp}}\int d V \ d W \left(F_0+\epsilon_{\delta} F_1\right)\mathcal J_0^{\mathrm{gc}} \left( \phi_1\right).
\end{eqnarray}
Finally, taking into account the expression for the second order Hamiltonian $H_2^{\mathsf{ORB5}}$,
we get:
\begin{eqnarray}
\nonumber
\mathcal{E}^{\mathsf{ORB5}}&=&
\sum_{\mathrm{sp}} \int dV\ dW\ 
\left(H_0-\epsilon_{\delta}\ e \frac{p_z}{m}\ \mathcal J_0 ^{\mathrm gc}\left(A_{1 \|} \right)\right) \ \left(F_0+\epsilon_{\delta} F_1\right)+
\sum_{\mathrm{sp}} \int dV \frac{\left|\bm\nabla_{\perp} A_{1\|}\right|^2}{8\pi} 
\\
\nonumber
&+&\sum_{\mathrm{sp}} \int dV\ dW \left(\frac{e^2}{2 mc^2} A_{1\|}^2+\frac{mc^2}{2 B^2}\left|\bm\nabla_{\perp}\phi_1\right|^2\right) F_0
\equiv \mathcal{E}_k+\mathcal{E}_F,
\end{eqnarray}
which corresponds to the energy density obtained from the direct application of the Noether method in Eulerian variational framework with the truncated Hamiltonian corresponding to the ORB5 model.
We refer to the first term, which contains only the unperturbed Hamiltonian $H_0$ as a kinetic energy $\mathcal{E}_k$ and the other terms as a field energy $\mathcal{E}_F$.

We proceed with the derivation of the code diagnostics. What can be measured in the PIC code in order to control the quality of the simulations?
It is well known that in PIC codes, particles and fields are evaluated in two different ways. Particles are advanced along their characteristics without use of any grid, while fields are evaluated on a grid.
So, to control the quality of the simulation, the contributions from the energy of the particles and the energy of the field should be computed independently.
This is why we are considering the power balance equation, called also the $E\times B$ transfer equation:
\begin{equation}
\frac{d{\mathcal E}_k}{d t}=-\frac{d\mathcal E_F}{d t}.
\label{energy_balance}
\end{equation}
The contributions from the particles are contained in the kinetic part ${\mathcal E}_k$ of the conserved energy density $\mathcal E$:
\begin{equation*}
\frac{d {\mathcal E}_k}{d t} =\sum_{\mathrm{sp}}\ \int dW\ dV\ \frac{d H_0}{d t}\ \left(F_0+\epsilon_{\delta}F_1\right) 
+\sum_{\mathrm{sp}} \int dW\ dV\ H_0\underbrace{\frac{d F}{d t}}_{\equiv 0}.
\end{equation*}
The first term here represents the explicit time derivative of the guiding-center Hamiltonian $H_0$.
The last term vanishes because of the Liouville theorem, and the remaining terms do not contain any dynamical fields.

To derive the diagnostics for the field part of the energy, as it is measured in the simulations, we need to use both the corresponding quasi-neutrality and the Amp\`ere equations, but this time replacing polarization terms related to the background distribution $F_0$ by the moments of the distribution function $F_0+\epsilon_{\delta}F_1$.

We start by writing the quasi-neutrality equation in a \textit{weak} form, taking into account additional integration by parts in order to replace the guiding-center gyro-averaged operator $\mathcal J_0^{\mathrm{gc}}$ from the electrostatic potential to the distribution function $F_0+\epsilon_{\delta}F_1$:
\begin{eqnarray*}
\sum_{\mathrm{sp}}\int dV dW \left(\frac{m c^2}{B^2} F_0\right) \left|\bm\nabla_{\perp}\phi_1\right|^2=
\sum_{\mathrm{sp}}\int dV dW \mathcal J_0^{\mathrm{gc}\dagger}\left(F_0+\epsilon_{\delta}F_1\right)\ \phi_1.
\end{eqnarray*}
Next, we reconstruct the Amp\`ere's equation corresponding to the ORB5 code in a \textit{weak}  form by combining the field terms in the energy density expression (\ref{ORB5_Energy}):
\begin{eqnarray*}
-\int dV\ dW\ F_0\ \frac{e^2}{2 m c^2}A_{1\|}^2+\int dV\ dW\ \frac{|\bm\nabla_{\perp} A_{1\|}|^2}{8\pi}=
\frac{1}{2}\int dV\ dW\ \frac{e p_z}{m c}\mathcal J_0 ^{\mathrm{gc}\dagger}\ \left(F_0+\epsilon_{\delta}F_1\right)   A_{1\|}.
\end{eqnarray*}
That operation leads to the following expression for the field energy term associated with the second order reduced dynamics, implemented in the energy balance equation (\ref{energy_balance}) :
\begin{eqnarray*}
\mathcal{E}_F=\sum_{\mathrm{sp}}\frac{1}{2}\int dV\ dW\ \mathcal J_0^{\mathrm{gc}\dagger} \left(F_0+\epsilon_{\delta} F_1\right)\ \phi_1-
\sum_{\mathrm{sp}}\frac{1}{2}\int dV\ dW\ \mathcal J_0^{\mathrm{gc}\dagger} \left(F_0+\epsilon_{\delta} F_1\right)\ \frac{e p_z}{mc}\ A_{1\|}.
\end{eqnarray*}
Therefore, the final expression for the energy density $\mathcal E$ is rewritten as:
\begin{eqnarray*}
\mathcal{E}=\sum_{\mathrm{sp}} \int dV dW \left(F_0+\epsilon_{\delta}F_1\right)
\left[\left(\frac{p_z^2}{2 m}+\mu B\right)+\frac{e}{2}\mathcal J_0^{\mathrm gc}\left(\phi_1\right)-\frac{p_z}{2 m c} \mathcal J_0^{\mathrm gc} \left(A_{1\|}\right)\right].
\end{eqnarray*}
We evaluate the time derivative of the kinetic energy $\mathcal E_k$ using the first order gyrocenter characteristics for
 the phase space coordinates  $\dot{p}_z$ and 
$\dot{\mathbf X}$:
\begin{eqnarray}
&&\frac{d\mathcal E_k}{d t} =\sum_{\mathrm{sp}} \int dV dW \left(F_0+\epsilon_{\delta}F_1\right) 
\left[\frac{p_z}{m}\ \dot{p}_{z}+
\mu\ \dot{\mathbf X}\cdot\bm\nabla B\right]
\nonumber
\\
&=&
\sum_{\mathrm{sp}} 
\int dV dW \left(F_0+\epsilon_{\delta}F_1\right)\ \left[-\left.e \bm\nabla\mathcal J_0^{\mathrm{gc}} \left(\psi_1\right)\cdot\dot{\mathbf{X}}\right|_0+
\left.\frac{1}{c} \mathcal J_0^{\mathrm{gc}}\left(A_{1\|}\right)\left(\frac{\dot{p}_z}{m}\right)\right|_0\right].
\label{J.E}
\end{eqnarray}
The details of that calculation are found in Appendix \ref{app:ORB5_diagnostics}.
For practical reasons, in numerical simulations, it is particularly useful to consider the power balance equation in the following form (i.e., normalized by the field energy $\mathcal E_F$):
\begin{eqnarray}
\frac{1}{\mathcal E_F}\frac{d\mathcal E_k}{d t}=-\frac{1}{\mathcal E_F}\frac{d\mathcal E_F}{d t}.
\label{energy_balance_norm}
\end{eqnarray}

\begin{figure}[h]
\begin{center}
\includegraphics[height=8cm,width=10cm]{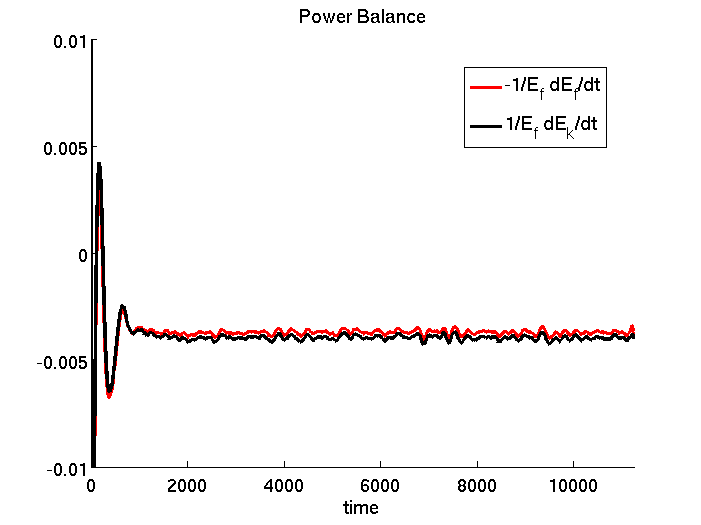}
\caption{Time evolution of the right-hand side and the left hand side of the power balance equation (\ref{energy_balance_norm}) for the linear CYCLONE base case simulations with the ORB5 code.}
\label{polarization}
\end{center}
\end{figure}

\begin{figure}[h]
\begin{center}
\includegraphics[height=8cm,width=10cm]{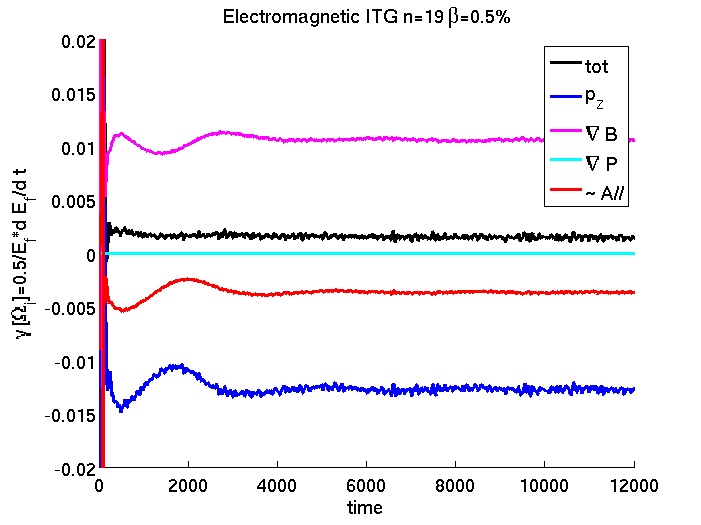}
\caption{Ion Temperature Gradient instability: time evolution of the different contributions to the instantaneous growth rate for the most unstable mode of the linear CYCLONE base case.}
\label{polarization1}
\end{center}
\end{figure}

\begin{figure}[h]
\begin{center}
\includegraphics[height=8cm,width=10cm]{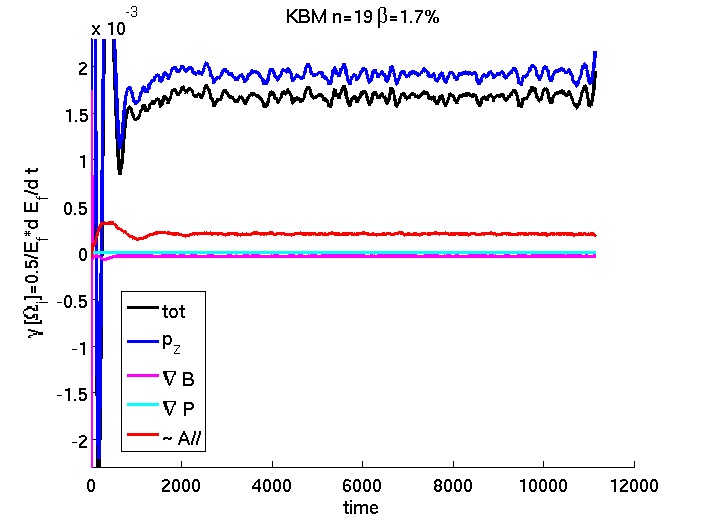}
\caption{Kinetic Ballooning Mode instability: time evolution of the different contributions to the instantaneous growth rate for the most unstable mode of the linear CYCLONE base case.}
\label{polarization2}
\end{center}
\end{figure}

In linear simulations, the power balance equation (\ref{energy_balance_norm}) not only gives an indication about the quality of the simulation 
but also can be used to measure the instantaneous growth rate of the instability:
$$
\gamma =\frac{1}{2}\frac{d}{dt} \log{\mathcal E}_F=\frac{1}{2}\frac{1}{\mathcal E_F}\frac{d{\mathcal E}_F}{dt}.
$$
Taking into account Eq. (\ref{J.E}), $\gamma$ becomes:
\begin{eqnarray*}
\gamma =\frac{1}{2\mathcal E_F}\sum_{\mathrm{sp}}\int dV\ dW\ \left(F_0+\epsilon_{\delta}F_1\right)
 \left.\left[e \bm\nabla \left\langle\psi_{1 \mathrm{gc}}\right\rangle\cdot\dot{\mathbf{X}}\right |_0-
\left.\frac{1}{c} \left\langle A_{1\| \mathrm{gc}}\right\rangle\left(\frac{\dot{p}_z}{m}\right)\right |_0\right].
\end{eqnarray*}
We represent on Figs.~\ref{polarization1} and \ref{polarization2}, examples of the diagnostics implementation for different types of instabilities, Electromagnetic Ion Temperature Gradient (ITG) and Kinetic Ballooning Mode (KBM).
The different contributions to the growth rate $\gamma$ arising from the different terms in the unperturbed guiding-center characteristics 
$\dot{\mathbf X}|_0$ and $\dot{p}_z |_0$ can be separated in the power balance equation and give a clear vision of which type of instability is present in the system: this diagnostics is suitable for both linear and nonlinear simulations:
\begin{eqnarray*}
\gamma
&=& 
\frac{1}{2\mathcal E_F}\sum_{\mathrm{sp}}\int dV \ dW \left(F_0+\epsilon_{\delta}F_1\right)
 \bm\nabla {\mathcal J_0}^{\mathrm{gc}}\left(\psi_1\right)\cdot
\left(\mathbf{v}_{\|}+\mathbf{v}_{{\bm\nabla} P}+\mathbf{v}_{{\bm\nabla} B}\right)
\nonumber
\\
&&-\frac{1}{2\mathcal E_F}\sum_{\mathrm{sp}}\int dV \ dW \left(F_0+\epsilon_{\delta}F_1\right) {\mathcal J_0}^{\mathrm{gc}}\left(A_{1\|}\right)
\left(
\mu B \bm\nabla\cdot\widehat{\mathbf b} + \frac{\mu c}{e B_{\|}^*} p_z \widehat{\mathbf b}\times
\left(\widehat{\mathbf b}\times\frac{\bm\nabla\times\mathbf{B}}{B}\right)\cdot\bm\nabla B\right),
\nonumber
\end{eqnarray*}
where
\begin{equation*}
\mathbf{v}_{\|}\equiv\frac{p_z}{m}\widehat{\mathbf b},
\end{equation*}
\begin{equation*}
\mathbf{v}_{\bm{\nabla}P}\equiv-\left(\frac{p_z}{m}\right)^2\frac{m c}{e B_{\|}^*}\widehat{\mathbf b}\times\frac{\bm\nabla P}{B^2},
\end{equation*}
and
\begin{equation*}
\mathbf{v}_{\bm{\nabla}B}\equiv \left(\frac{\mu B}{m}+
\left(\frac{p_z}{m}\right)^2\right)\frac{m c}{e B_{\|}^*}\widehat{\mathbf b}\times\frac{\bm\nabla B}{B}.
\end{equation*}

%
\section{Conclusions and perspectives}
%
In this work we have presented two variational derivations of self-consistent gyrokinetically reduced systems of the Maxwell-Vlasov equations containing the second order corrections with respect to the small parameter related to the gyrocenter dynamical reduction $\epsilon_{\delta}$ and up to the second order with respect to the FLR corrections. The first system issued from the direct derivation uses second order truncated Eulerian variational action functional for the gyrokinetic Maxwell-Vlasov system. The second system uses some physical approximations, suitable for PIC code implementations.

The main results of this work can be summarized in the following way: The electrostatic limit of both models coincides. The electrostatic model implemented in ORB5 is energetically consistent.
Concerning the electromagnetic case, several differences exist: First, due to the differences in the second order Hamiltonians, the gyrokinetic equations for the electromagnetic fields may or may not be coupled.
Also the terms proportional to $\nabla_{\perp}^2 A_{1\|}$ have different expressions.
Neglecting the magnetic field curvature in the model issued from the physical approximation violates the constraint $\nabla\cdot{\mathbf B}=0$, which can be a potential issue while implementing a phase space preserving numerical scheme.
Finally, the reduced Vlasov equations coincide up to the nonlinear term $\{F_1,H_1\}_{\mathrm{gc}}$, issued from the direct second order derivation, which is absent in a physical model.

At the next step of our work we aim to proceed with the derivation of the weak form for Eqs.~(\ref{Poisson_full}) and (\ref{Ampere_full}) suitable for the finite element method discretization, necessary for the implementation of the coupled system in ORB5. In addition, further comparison of the established second order reference model (\ref{Poisson_full}, \ref{Ampere_full}) with the gyrokinetic equations implemented into other codes involved into the Enabling Research project is one of our highest priorities.

\section{Acknowledgments}

Authors would like to thank A.J. Brizard and B.D. Scott for helpful discussions. The simulations have been performed on Helios CSC supercomputer in the framework of VeriGyro and ORBFAST projects. Authors also thank the Referee for very detailed revision, useful, constructive suggestions and remarks.

\textit{This work has been carried out within the framework of the EUROfusion Consortium and has received funding from the Euratom research and training programme 2014-2018 under grant agreement No 633053. The views and opinions expressed herein do not necessarily reflect those of the European Commission.}
%
\appendix
%
%
\section{\label{app: rho1}First order gyrocenter displacement}
%

The gyrocenter displacement $\bm\rho_1$ appears as the shift between the position of the particle and the position of the guiding-center. It is obtained by a Lie transform with a generating function $S_1$ to be determined:
\begin{eqnarray*}
e^{-\pounds_{S_1}}\left(\mathbf X+\bm\rho_0\right) =\left(\mathbf X+\bm\rho_0\right)-\left\{S_1,\mathbf X+\bm\rho_0\right\}
+\mathcal{O}\left(\epsilon_{\delta}^2\right)=
\mathbf X+\bm\rho_0+\bm\rho_1.
\end{eqnarray*}
Therefore, at the first order in $\epsilon_{\delta}$, the expression for the lowest order gyrocenter displacement is given by
\begin{equation}
\bm\rho_1=-\left\{S_1, \mathbf{X}+\bm\rho_0\right\}_{\mathrm{gc}},
\label{rho_gy_1_impl}
\end{equation}
where $S_1$ is the lowest order generating function of the gyrocenter transformation. 
The generating function of the Lie transform is determined such that it eliminates the fluctuating part depending on the fast variables, namely, the gyroangle. The generating function $S_1$ eliminates the order $\epsilon_\delta$ and $S_2$ eliminates the second order of $H_{\rm gy}=H_{\rm gc}+\epsilon_\delta H_1+\epsilon_\delta^2 H_2$ given by Eq.~(\ref{eqn:Hgy}). The Hamiltonian expressed in the new variables becomes: 
\begin{eqnarray*}
\overline{H}_{\mathrm{gy}}
&=&e^{-\pounds_{S_2}}e^{-\pounds_{S_1}}
H_{\mathrm{gy}}=
H_{\mathrm{gc}}-\left\{S_1, H_{\mathrm{gc}}\right\}+\frac{1}{2}\left\{S_1,\left\{S_1, H_{\mathrm{gc}}\right\}\right\}-
\left\{S_2, H_{\mathrm{gc}}\right\}
\nonumber
\\&+&
e\ \psi_{1}(\mathbf{X}+\bm\rho_0)+e\bm\rho_1\cdot\bm\nabla\psi_{1}(\mathbf{X}+\bm\rho_0)+
\frac{1}{2 m}\left(\frac{e}{c}\right)^2A_{1\|}(\mathbf{X}+\bm\rho_0)^2+\mathcal{O}\left(\epsilon_{\delta}^3\right).
\end{eqnarray*}
In the above expression, we have used the fact that
$$
e^{-\pounds_{S_1}} \psi_1(\mathbf{X}+\bm\rho_0)=\psi_1(\mathbf X+\bm\rho_0+\bm\rho_1),
$$
and the expansion
$$
\psi_1(\mathbf X+\bm\rho_0+\bm\rho_1)=\psi_1(\mathbf X+\bm\rho_0)+\bm\rho_1\cdot\bm\nabla\psi_{1}(\mathbf{X}+\bm\rho_0)+\mathcal{O}(\epsilon_{\delta}^3). 
$$
The expression for $S_1$ is obtained from the condition that the gyrophase dependent part of the linear electromagnetic perturbation 
$\widetilde{\psi}_1$
is removed from the lowest order gyrocenter Hamiltonian:
\begin{equation}
\left\{S_1,H_{\mathrm{gc}}\right\}=e\  \widetilde{\psi}_1\left(\mathbf{X}+\bm\rho_0\right)=e  \psi_1\left(\mathbf{X}+\bm\rho_0\right)-e \langle  {\psi}_1\left(\mathbf{X}+\bm\rho_0\right)\rangle.
\label{S_1_bracket}
\end{equation}
By taking into account the expression for the guiding-center Poisson bracket, the above condition becomes:
\begin{equation*}
\frac{e}{m c}\ \frac{\partial S_1}{\partial\theta}\ \frac{\partial H_{\mathrm{gc}}}{\partial\mu}=
\frac{e B}{m c}\ \frac{\partial S_1}{\partial\theta}=e\ \widetilde{\psi}_1\left(\mathbf{X}+\bm\rho_0\right),
\end{equation*}
and therefore the generating function is given by:
$
S_1=\frac{e}{\Omega}\int  d\theta\  \widetilde{\psi}_1\left(\mathbf{X}+\bm\rho_0\right).
$
At the lowest guiding-center order, $\widetilde{\psi}_1\left(\mathbf{X}+\bm\rho_0\right)=\rho_0\widehat{\bm\rho}\cdot\bm\nabla\psi_1(\mathbf{X})$, and consequently
\begin{equation*}
S_1=\frac{mc}{B}\rho_0\ \widehat{\perp}\cdot\bm\nabla\psi_1(\mathbf{X}),
\end{equation*}
where we have used the property of the rotating basis vectors  $\widehat{\bm\rho}=\int d\theta\ \widehat{\perp}$.

Now we use the expression for the generating function $S_1$, which contains to the lowest order guiding-center corrections to calculate the order $\epsilon_{\delta}$ of the gyrocenter displacement (\ref{rho_gy_1_impl}).
Using the expression for the guiding-center Poisson bracket:
\begin{eqnarray*}
\bm\rho_1=-\frac{e}{mc}\left(\frac{\partial S_1}{\partial\theta}\frac{\partial\bm\rho_0}{\partial\mu}-\frac{\partial\bm\rho_0}{\partial\theta}\frac{\partial S_1}{\partial\mu}\right).
\end{eqnarray*}
From Eq.~(\ref{rho_0}), we have $\partial_{\mu}\rho_0^2=\frac{2 m c^2}{e^2 B}$, we have:
\begin{eqnarray*}
\frac{e}{mc}\frac{\partial S_1}{\partial\theta}\frac{\partial\bm\rho_0}{\partial\mu}&=&\frac{m c^2}{e B^2}\ \widehat{\bm\rho}\widehat{\bm\rho}\cdot\bm\nabla\psi_1\\
-\frac{e}{mc}\frac{\partial\bm\rho_0}{\partial\theta}\frac{\partial S_1}{\partial\mu}&=&\frac{m c^2}{e B^2}\ \widehat{\perp}\widehat{\perp}\cdot\bm\nabla\psi_1
\end{eqnarray*}
By taking into account the definition of the dyadic tensor $\bf{1}_{\perp}\equiv\widehat{\bm\rho}\widehat{\bm\rho}+\widehat{\perp}\widehat{\perp}$, the expression for the first order gyrocenter displacement in the \textit{long wavelength approximation} is:
\begin{equation}
\bm\rho_1=-\frac{m c^2}{e B^2}\bm\nabla_{\perp}\psi_1.
\end{equation}
This demonstrates the link between the definition of the reduced particle position, and in particular the displacement ${\bm\rho}_1$, and the elimination of the gyrophase dependence of the reduced Hamiltonian dynamics.

The generating function $S_2$ is defined such that it removes the gyroangle dependence from the order $\epsilon_{\delta}^2$ terms. Hamiltonian $\overline{H}_{\rm gy}$ becomes
\begin{eqnarray*}
\overline{H}_{\mathrm{gy}}
&=&
H_{\mathrm{gc}}+e \langle \psi_{1}(\mathbf{X}+\bm\rho_0)\rangle+\frac{e}{2}\left\langle \left\{S_1,\widetilde{\psi}_1\left(\mathbf{X}+\bm\rho_0\right) \right\}\right\rangle \\
&& +e\langle \bm\rho_1\cdot\bm\nabla\psi_{1}(\mathbf{X}+\bm\rho_0)\rangle+
\frac{1}{2 m}\left(\frac{e}{c}\right)^2\langle A_{1\|}(\mathbf{X}+\bm\rho_0)^2\rangle+\mathcal{O}\left(\epsilon_{\delta}^3\right).
\end{eqnarray*}
At the leading order in $\rho_0$, from the definition of the generating function $S_1$, we have
$$
\left\langle \left\{S_1,\widetilde{\psi}_1\left(\mathbf{X}+\bm\rho_0\right) \right\}\right\rangle = -\langle\bm\rho_1 \cdot \bm\nabla\psi_1\rangle.
$$
Therefore, using the above expression for $\bm\rho_1$, the second order gyrocenter Hamiltonian becomes
$$
\overline{H}_{\mathrm{gy}}=
H_{\mathrm{gc}}+e \langle \psi_{1}(\mathbf{X}+\bm\rho_0)\rangle -\frac{mc^2}{2B^2}\vert \bm\nabla_\perp \psi_1 \vert^2+
\frac{1}{2 m}\left(\frac{e}{c}\right)^2\langle A_{1\|}(\mathbf{X}+\bm\rho_0)^2\rangle.
$$
%
\section{\label{app:ORB5_diagnostics}Hamiltonian first order characteristics and ORB5 code diagnostics}
%

In that Appendix we give a derivation of the first order gyrocenter characteristics in the Hamiltonian representation.
This will allow us to explicit the diagnostics implemented in the ORB5 code for the control of the quality of simulations.

In the $p_z$ representation, the magnetic field $\mathbf B^*$ writes:
\begin{equation*}
{\mathbf B}^*={\mathbf B}+\frac{c}{e}\ p_z\ \bm\nabla\times\widehat{\mathbf b}.
\end{equation*} 

The geometric contribution to this symplectic field $\bm\nabla\times\widehat{\mathbf  b}$ is expressed with using the projection on the parallel and perpendicular directions:
\begin{eqnarray*}
\bm\nabla\times\widehat{\mathbf b}=\widehat{\mathbf b}\left(\widehat{\mathbf b}\cdot\bm\nabla\times\widehat{\mathbf b}\right)-
\widehat{\mathbf b}\times\left[\widehat{\mathbf b}\times\bm\nabla\times\widehat{\mathbf b}\right]\equiv \widehat{\mathbf b}\ \tau- {\mathbf G},
\end{eqnarray*}
where the scalar $\tau$ represents the magnetic twist and the vector $\mathbf G$ is referred to as the magnetic curvature.
Since $\mathbf B\times \left(\bm\nabla\times{\mathbf B}\right)= -\nabla p$ in single fluid MHD equilibrium, 
we rewrite the curvature vector $\mathbf G$ in the following way in order to evidence the pressure-like contributions into the characteristics:
\begin{eqnarray*}
\mathbf G=\widehat{\mathbf b}\times\left[\widehat{\mathbf b}\times\frac{\bm\nabla\times{\mathbf B}}{B}\right] +
\frac{\bm\nabla B\times\widehat{\mathbf b}}{B}.
\end{eqnarray*}
We also decompose the symplectic magnetic field in the parallel and perpendicular components in the following way:
\begin{eqnarray*}
\mathbf B^*= \underbrace{\left(B +\frac{c}{e} p_z \ \tau \right)}_{\equiv B_{\|}^*}\widehat{\mathbf b} -\frac{c}{e} p_z \mathbf{G}.
\end{eqnarray*}
The final expressions for the characteristics are implemented in the code in the following form:
\begin{eqnarray}
\dot{\mathbf X}&=&\frac{p_z}{m}\ \widehat{\mathbf b}-\left(\frac{p_z}{m}\right)^2\frac{m}{e B_{\|}^*}\widehat{\mathbf b}
\times\left(\widehat{\mathbf b}\times\frac{\bm\nabla\times{\mathbf B}}{B}\right)
+\left(\frac{\mu}{m}+\left(\frac{p_z}{m}\right)^2\right)\frac{m}{e B_{\|}^*}
\widehat{\mathbf b}\times\frac{\bm\nabla B}{B} 
\nonumber
\\
&-&\frac{e}{c}\left\langle A_{1\| \mathrm{gc}}\right\rangle \widehat{\mathbf b}
\label{R_0}
+\frac{p_z}{m}\frac{1}{B_{\|}^*}
\left\langle A_{1\| \mathrm{gc}}\right\rangle\mathbf G- \frac{1}{B_{\|}^*} \bm\nabla \left\langle\psi_{\mathrm{1 \mathrm{gc}}}\right\rangle\times \widehat{\mathbf b},
\label{elm_pert}
\\
&\equiv& \mathtt{vpar}+\mathtt{vpressure}+\mathtt{vgradb}+\mathtt{vapar1}+\mathtt{vexb}.
\end{eqnarray}

The first three terms represent the non-perturbed (guiding-center) characteristics with $\mathtt{vpar}$ the parallel velocity, 
$\mathtt{vpressure}$ the pressure-like term and 
$\mathtt{vgradb}$ containing the gradient of magnetic field amplitude $\bm\nabla B$. The next two terms contain the perturbed gyrocenter 
electromagnetic potential $\left\langle A_{1\| \mathrm{gc}}\right\rangle$ and are referred to as $\mathtt{vapar1}$. The last term $\mathtt{vexb}$
is the electromagnetic ${\bf E}\times {\bf B}$ velocity.

The characteristics for $p_z$ are given by
\begin{eqnarray}
\dot{p}_z &=& \mu B\ \bm\nabla\cdot\widehat{\mathbf b} + \frac{\mu c}{e B_{\|}^*} p_z \widehat{\mathbf b}\times
\left(\widehat{\mathbf b}\times\frac{\bm\nabla\times\mathbf{B}}{B}\right)\cdot\bm\nabla B
\label{GC_pz}
\\
&-&e \bm\nabla \left\langle\psi_{\mathrm{1 gc}}\right\rangle\cdot
\left(\widehat{\mathbf b}-\frac{c}{e B_{\|}^*} p_z\mathbf G\right),
\label{Pert_pz}
\\
&\equiv&
{\mathtt{dvapdt0}}+{\mathtt{dvapdt1}},
\end{eqnarray}
where we have used the divergence free property of magnetic field: $\widehat{\mathbf b}\cdot\bm\nabla B=- B\ \bm\nabla\cdot\widehat{\mathbf b}$.
We have also organized the terms in two groups: the unperturbed guiding-center contributions $\mathtt{dvapdt0}$ and those containing linear gyro-averaged electromagnetic potential $\mathtt{dvapdt1}$:
\begin{equation}
\left\langle\psi_{\mathrm{1 \mathrm{gc}}}\right\rangle=\left\langle\phi_{\mathrm{1 gc}}\right\rangle-\frac{1}{c}\ \frac{p_z}{m} \ \left\langle A_{1\| \mathrm{gc}}\right\rangle.
\end{equation}
%
\section{\label{app:Noether_method} Application of Noether's method to gyrokinetics}
%
In this Appendix we sketch the main steps of the Noether's method for the systematic derivation of conservation laws.
We start from the Noether equation:
\begin{eqnarray*}
\frac{\partial{\mathsf S}}{\partial t}+\bm\nabla\cdot{\mathbf J}=\delta{\mathcal L}^{\mathcal E}.
\end{eqnarray*}
First, from Sec.~\ref{sec:IIIA}, we collect all the exact derivative terms (Noether terms) 
\begin{eqnarray*}
0&=&
-\int d^8 Z \left\{\widehat{\cal S}{\mathcal H}_1,{\mathcal F} \right\}_{\mathrm{ext}}
\nonumber
\\
&-&\epsilon_\delta^2 \int \frac{dV dt}{4\pi}\bm{\nabla}\cdot\left({\mathbf E}_1\ \widehat\phi_1\right)
-\epsilon_\delta^2 \int\frac{dV dt}{4\pi}\frac{1}{c}\ \partial_t\left({\mathbf E}_1\ \cdot\widehat {\mathbf A}_1\right)
\\
\nonumber
&+&
\epsilon_\delta \int \frac{dV dt}{4 \pi}\bm\nabla\cdot\left(\left({\mathbf B}_0+\epsilon_\delta {\mathbf B}_1\right)\times\widehat{\mathbf A}_1\right).
\end{eqnarray*}
The main idea is to separate the temporal derivatives (density terms) from the spatial derivatives (flux terms). As we can see, the three last terms of the above equation, obtained from the Maxwell part of the action functional, can already be identified as flux and density terms.
The terms obtained from the Vlasov part of the Eulerian action functional require some additional manipulations.
First we explicitly write the expression for the Poisson bracket:
\begin{eqnarray}
&-&\int d^8 Z \left\{\widehat{\cal S}{\mathcal H}_1,{\mathcal F} \right\}_{\mathrm{ext}}=\int d^8 Z \left\{{\mathcal F},\widehat{\cal S}{\mathcal H}_1 \right\}_{\mathrm{ext}}
\equiv\int d^8 Z
\left(\frac{\partial}{\partial Z^a} {\mathcal F}\right) J^{a b} \frac{\partial}{\partial Z^b}\left( \widehat{\cal S}{\mathcal H}_1\right)
\nonumber 
\\
&=& 
\int d^8 Z\  J^{a b}  \  \frac{\partial}{\partial Z^a}\left( {\mathcal F} \frac{\partial}{\partial Z^b}\left(\widehat{\cal S}{\mathcal H}_1\right)\right)
-
\int d^8 Z \  J^{a b}  \ \ {\mathcal F}\ \frac{\partial^2}{\partial Z^a\partial Z^b}\left(\widehat{\cal S}{\mathcal H}_1 \right), 
\label{leibnitz1}
\end{eqnarray}
where $J^{ab}$ denotes the Poisson matrix. Since it is antisymmetric, the last term vanishes.

Next we are using Liouville's theorem for Hamiltonian $\widehat{\cal S}{\mathcal H}_1$: $\bm\nabla\cdot\dot{\bf Z}=0$, with $J= \det\left|J^{ab}\right|$ is the determinant of the Poisson matrix, and $\dot{Z}^a=J^{a b}\frac{\partial (\widehat{\cal S}{\mathcal H}_1)}{\partial Z^b}$. We notice the following identity:
\begin{eqnarray*}
0=\frac{1}{J} \frac{\partial}{\partial Z^a}\left(J \dot{Z}^a\right)= \frac{1}{J} \frac{\partial}{\partial Z^a}\left(J J^{a b}\frac{\partial (\widehat{\cal S}{\mathcal H}_1)}{\partial Z^b}\right)=
\underbrace{J^{a b}\frac{\partial^2 (\widehat{\cal S}{\mathcal H}_1)}{\partial Z^a \partial Z^b}}_{*}
+\underbrace{\frac{1}{J}\frac{\partial}{\partial Z^a}\left(J  J^{a b}\right)\frac{\partial (\widehat{\cal S}{\mathcal H}_1)}{\partial Z^b}}_{**},
\end{eqnarray*}
where the term $*$ vanishes because of the antisymmetry of the Poisson matrix. 
We rewrite the non-zero term of Eq.~(\ref{leibnitz1}):
\begin{eqnarray*}
\int d^8 Z\  J^{a b}  \  \frac{\partial}{\partial Z^a}\left( {\mathcal F} \frac{\partial (\widehat{\cal S}{\mathcal H}_1)}{\partial Z^b}\right)&=&
\int d^8 Z\  \frac{1}{J}  \frac{\partial}{\partial Z^a}\left(
J {\mathcal F}
\underbrace{J^{a b}\ \frac{\partial (\widehat{\cal S}{\mathcal H}_1)}{\partial Z^b}}_{\equiv\left\{Z^a,\widehat{\cal S}\mathcal{H}_1\right\}_{\mathrm{ext}}}\right)
\\
&-&
\int d^8 Z\ {\mathcal{F}}\ \underbrace{\frac{1}{J} \ \frac{\partial}{\partial Z^a}\left(J J^{a b}\right)
\frac{\partial (\widehat{\cal S}{\mathcal H}_1)}{\partial Z^b}}_{=0},
\nonumber
\end{eqnarray*}
where the last term vanishes due to Liouville's theorem.
Therefore, we have rewritten the Noether's contribution as follows:
\begin{eqnarray*}
-\int d^8 Z \left\{\widehat{\cal S}{\mathcal H}_1,{\mathcal F} \right\}_{\mathrm{ext}}=
\int d^8 Z \frac{1}{J}\frac{\partial}{\partial Z^a}\left( J {\mathcal F} \left\{Z^a, {\widehat{\cal S}\mathcal{H}_1}\right\}_{\mathrm{ext}}\right).
\end{eqnarray*}
We are now writing the explicit expression for the phase-space volume element $d^8 Z\equiv J\ d^4 x\ d^4 p\equiv J\ d^3 X dt \ d^3 p\ dw$ and with introducing the four-vectors for the energy-momentum  $p^{\nu}\equiv (w,p^i)$ and space-time $x^{\mu}\equiv(ct, X^j)$:
\begin{eqnarray*}
&&\int d^8 Z \frac{1}{J}\frac{\partial}{\partial Z^a}\left( J {\mathcal F} \left\{Z^a, {\widehat{\cal S}\mathcal {H}_1}\right\}_{\mathrm{ext}}\right)=
\int d^4 x \underbrace{\int d^4 p \frac{\partial}{\partial p^{\nu}}\ \left( J {\mathcal F} \left\{p^{\nu}, {\widehat{\cal S}\mathcal{H}_1}\right\}_{\mathrm{ext}}\right)}_
{=0}
\nonumber
\\
&+&
\int d^4 x \int d^4 p \frac{\partial}{\partial x^{\mu}} \left( J  {\mathcal F} \left\{x^{\mu}, {\widehat{\cal S}\mathcal {H}_1}\right\}_{\mathrm{ext}}\right)
=
\int d^4 x \frac{\partial}{\partial x^{\mu}} \int d^4 p\ \left( J  {\mathcal F} \left\{x^{\mu}, {\widehat{\cal S}\mathcal{H}_1}\right\}_{\mathrm{ext}}\right). 
\nonumber
\end{eqnarray*}
Here the term with the energy-momentum derivatives vanishes since it is an exact derivative. The term which contains the spatial derivatives can be rewritten, taking the spatial derivative out of the integral.
This procedure allows the separation of the Noether density and flux contributions obtained from the Vlasov terms.
Taking into account that
\begin{eqnarray*}
\left\{x^{\mu}, {\widehat{\cal S}\mathcal {H}_1}\right\}_{\mathrm{ext}}=\left\{x^{\mu}, {\widehat{\cal S}}\right\}_{\mathrm{ext}}\mathcal{H}_1+\left\{x^{\mu}, {\mathcal{H}}_1\right\}_{\mathrm{ext}}\widehat{\cal S}, 
\end{eqnarray*}
and using the definition of the extended Vlasov field $ {\mathcal F}\equiv\  {F}\ \delta(w-H_0-\epsilon_\delta H_1)$ and the Hamiltonian ${\mathcal H}_1\equiv H_0+\epsilon_\delta H_1-w$, we have
\begin{eqnarray*}
\int d w\  {\mathcal F} \left\{x^{\mu},\widehat{\cal S}\right\}_{\mathrm{ext}}{\mathcal H}_1\equiv
\int d w\ \delta(w-H_0-\epsilon_\delta H_1)  {F} \left\{x^{\mu},\widehat{\cal S}\right\}_{\mathrm{ext}} (H_0+\epsilon_\delta H_1-w)=0.
\end{eqnarray*}
Finally, we have
\begin{eqnarray*}
\frac{\partial}{\partial x^{\mu}}\int d^4 p\  {\mathcal F} \left\{x^{\mu},{\mathcal H}_1\right\}_{\mathrm{ext}}\widehat{\cal S}=
\frac{1}{c} \frac{\partial}{\partial t}\int d^4 p\  {\mathcal F}\ \widehat{\cal S}\underbrace{\left\{c\ t,{\mathcal H}_1\right\}_{\mathrm{ext}}}_{\equiv c}+
\bm\nabla\cdot\int d^4 p\  {\mathcal F}\ \widehat{\cal S}\ \underbrace{\left\{{\bf X},{\mathcal H}_1\right\}_{\mathrm{ext}}}_{\equiv\dot{\bf X}}.
\end{eqnarray*}
Collecting the Noether density contributions from the Maxwell and the Vlasov part of the Eulerian action, we get:
\begin{eqnarray*}
{\mathsf S}\equiv-\frac{\epsilon_\delta^2}{4\pi c} {\mathbf E}_1\cdot\widehat{\mathbf A}_1 +\int d^4p\  {\mathcal F}\ \widehat{\cal S}.
\end{eqnarray*}
For the flux part we have:
\begin{eqnarray*}
{\mathbf J}=-\frac{\epsilon_\delta^2}{4\pi}{{\mathbf E}_1}\widehat{\phi}_1\ +
\frac{\epsilon_\delta}{4\pi}\left[\left({\mathbf B}_0+\epsilon_\delta{\mathbf B}_1\right) \times \widehat{\mathbf A}_1\right]
+
\int d^4p\  {\mathcal F}\ \widehat{\cal S}\{{\mathbf X},{{\cal H}_1}\}_{\rm ext}.
\end{eqnarray*}

%
\section{\label{app:Eulerian_explicit}Explicit derivation of the full second order Maxwell-Vlasov equations}
%

The first variation of the Eulerian action functional contains three parts, each one corresponds to the functional dependence in variational fields $(\phi_1, A_{1\|}, {\mathcal F})$:
\begin{eqnarray}
\delta{\mathcal A}^{\mathcal E}\left[\phi_1,A_{1\|},{\mathcal F}\right]=
\frac{\delta {\mathcal A}^{\mathcal E}}{\delta\phi_1}\circ\widehat{\phi}_1+
\frac{\delta {\mathcal A}^{\mathcal E}}{\delta A_{1\|}}\circ\widehat{A}_{1\|}+
\frac{\delta {\mathcal A}^{\mathcal E}}{\delta{\mathcal F}}\circ \delta\widehat{\mathcal F}.
\end{eqnarray}


%
\subsubsection{Fields contributions. Parallel magnetic perturbation}
%
Taking into account the expression for the perturbative magnetic field (\ref{B_1_par}), we compute the functional derivatives with respect to the parallel component of the magnetic potential $A_{1\|}$. We choose the test function $\widehat{\mathbf A}_1\equiv \widehat{A}_{1\|}\widehat{\mathbf b}$.
Following the implicit derivation presented in Sec.~\ref{subsec:implicit}, we start by calculating functional derivative with respect to $A_{1\|}$ [see also Eq.~(\ref{delta_field_magnetic_A_1})]. It leads to
\begin{eqnarray}
\frac{\delta{\mathcal A}^{{\mathcal E}, ({\mathsf{field}})}_{\mathsf{magn}}}{\delta {A}_{1\|}}\circ\widehat{A}_{1\|}&=&
-\epsilon_{\delta}\int\frac{dV}{4\pi}\left({\mathbf B}_0+\epsilon_{\delta}{\mathbf B}_1\right)\cdot\bm\nabla\times \left(\widehat{A}_{1\|}\widehat{\mathbf b}\right),
\nonumber \\
&=&-\epsilon_{\delta}\int\frac{dV}{4\pi}\  \widehat{A}_{1\|}\ \widehat{\mathbf b}\cdot\left(\bm\nabla\times\left[\mathbf{B}_0+\epsilon_{\delta}{\mathbf B}_1\right]\right)
\nonumber \\
&-&
\epsilon_{\delta}^2\int\frac{dV}{4\pi}\bm\nabla\cdot\left(\widehat{A}_{1\|}\left[\bm\nabla_{\perp}A_{1\|}+A_{1\|}\ \widehat{\mathbf b}\times(\bm\nabla\times\widehat{\mathbf b})\right]\right),
\label{eq:A1}
\end{eqnarray}
where for the Noether term [last line of Eq.~(\ref{eq:A1})] we have used the following identity:
$$
\left({\mathbf B}_0+\epsilon_{\delta}{\mathbf B}_1\right)\times\widehat{\mathbf A}_{1\|}
=
\epsilon_{\delta}\widehat{A}_{1\|}\left(\bm\nabla A_{1\|}\times\widehat{\mathbf b}\right)\times\widehat{\mathbf b}\
+\epsilon_{\delta}\widehat{A}_{1\|}A_{1\|}\left(\bm\nabla\times\widehat{\mathbf b}\right)\times\widehat{\mathbf b}.
$$
As we can see, the dynamical term [second line of Eq.~(\ref{eq:A1})] contains the unitary vector $\widehat{\mathbf b}$ of the background magnetic field, which means that the resulting Amp\`ere equation will be projected on the parallel direction. This is a direct consequence of the choice of the perturbed magnetic field (\ref{B_1_par}).

%
\subsubsection{\label{ssubsec:vlasov_first}First order Vlasov contributions}
%
The direct calculation from the second order Eulerian action gives us the following form of the first order Vlasov contributions [see also Eqs.(\ref{H1}) and (\ref{Vlasov_contr})]:
\begin{eqnarray*}
&& \frac{\delta{\mathcal A}_{\mathsf{lin}}^{{\mathcal E},{(\mathsf{Vl}})}}{\delta\phi_1}\circ\widehat{\phi}_1=
-e \epsilon_{\delta}\int \ dV\ dW \ \left(F_0+\epsilon_{\delta}F_1\right)\
 \left\langle\delta^3(\mathbf{X}+\bm\rho_0-{\mathbf r})\ \widehat{\phi}_1(\mathbf r)\right\rangle,
\\
&& \frac{\delta{\mathcal A}_{\mathsf{lin}}^{{\mathcal E},{(\mathsf{Vl}})}}{\delta A_{1\|}}\circ\widehat{A}_{1\|}=
e \epsilon_{\delta}\int \ dV\ dW \ \left(F_0+\epsilon_{\delta}F_1\right)\ 
\frac{p_z}{mc}\left\langle\delta^3(\mathbf{X}+\bm\rho_0-{\mathbf r})\ \widehat{A}_{1\|}(\mathbf r)\right\rangle, 
\end{eqnarray*}
where we have used:
\begin{eqnarray}
\frac{\delta H_1}{\delta\phi_1}\circ\widehat{\phi}_1&=&
e\frac{\delta\left\langle\phi_1\left({\mathbf X}+\bm\rho_0\right\rangle\right)}{\delta\phi_1(\mathbf r)}\circ\widehat{\phi}_1(\mathbf r)
\nonumber
\\
&=&
e\left\langle \delta^3({\mathbf X}+\bm\rho_0-\mathbf r)\widehat{\phi}_1(\mathbf r)\right\rangle=
e\left\langle \widehat{\phi}_1({\mathbf X}+\bm\rho_0) \right\rangle\equiv e \mathcal J_0^{\mathrm{gc}}\left(\widehat{\phi}_1\right), 
\label{delta_H_1_delta_phi_1}
\end{eqnarray}
and 
\begin{eqnarray}
\frac{\delta H_1}{\delta A_{1\|}}\circ\widehat{A}_{1\|}=-\frac{e p_z}{mc}\frac{\delta\left\langle  A_{1\|}\left({\mathbf X}+\bm\rho_0\right) \right\rangle}{\delta A_{1\|}}\circ\widehat{A}_{1\|}
=-\frac{e p_z}{m c} \left\langle \widehat{A}_{1\|}\left({\mathbf X}+\bm\rho_0\right) \right\rangle
\equiv-\frac{e p_z}{m c}  \mathcal J_0^{\mathrm{gc}}\left(\widehat{A}_{1\|}\right).
\label{delta_H_1_delta_A_1}
\end{eqnarray}

%
\subsubsection{\label{ssubsec:vlasov_second}Second order Vlasov contributions}
%
Next we explicitly compute the contributions from the nonlinear Vlasov terms associated with the second order Hamiltonian $H_2$ [Eq.~(\ref{H2_full_GK})]. We separate the latter in two parts: ${H}_{2 \mathsf{polmix}}$ and ${H}_{2 \mathsf{polmag}}$, which will generate two nonlinear Vlasov contributions to the nonlinear Vlasov part ${\mathcal A}^{\mathcal E,({\mathsf {Vl}})}_{\mathsf{nonlin}}$ of the Eulerian action defined in Eq.~(\ref{A_E_separation}):

\begin{equation}
H_{2 \mathsf{polmix}}=-
\frac{m c^2}{2 B^2} \left|\bm\nabla_{\perp}\phi_{1}-\frac{p_{z}}{m c}\bm\nabla_{\perp}A_{1\|}\right|^2,
\label{H_2polmix}
\end{equation}
and
\begin{equation}
H_{2 \mathsf{polmag}}=
\frac{e^2}{2 m c^2} A_{1\|}^2+\frac{\mu}{2 B} \left|\bm\nabla_{\perp}A_{1\|}\right|^2
+{\color{black}\frac{1}{2}\frac{\mu}{B}A_{1\|}\bm\nabla_{\perp}^2 A_{1\|}}.
\label{H_2polmag}
\end{equation}

The $H_{\mathsf {2 polmix}}$ part of the Hamiltonian will contribute to the electrostatic $\widehat{\phi}_1$ and magnetic $\widehat{A}_{1\|}$ parts of the first derivative of $H_2$, while ${H}_{2 \mathsf{polmag}}$ only contributes to the magnetic part:
\begin{eqnarray}
\frac{\delta H_{2\mathsf{polmix}}}{\delta \phi_{1}}\circ\widehat{\phi}_1=
-\frac{m c^2}{B^2}\left(\bm\nabla_{\perp}\phi_1-\frac{p_z}{mc}\bm\nabla_{\perp}A_{1\|}\right)\cdot\bm\nabla_{\perp}\widehat{\phi}_1,
\label{delta_H_2_delta_phi_1}
\end{eqnarray}
\begin{eqnarray}
\frac{\delta H_{2 \mathsf{polmix}}}{\delta A_{1\|}}\circ\widehat{A}_{1\|}=
\frac{m c^2}{B^2}\left(\bm\nabla_{\perp}\phi_1-\frac{p_z}{mc}\bm\nabla_{\perp}A_{1\|}\right)\cdot\left(\frac{p_z}{mc}
\bm\nabla_{\perp}\widehat{A}_{1\|}\right),
\label{delta_H_2_delta_A_1_polmix}
\end{eqnarray}
\begin{eqnarray}
\frac{\delta H_{2 \mathsf{polmag}}}{\delta A_{1\|}}\circ\widehat{A}_{1\|}=
\frac{e^2}{mc^2}A_{1\|}\widehat{A}_{1\|} +\frac{\mu}{B}\bm\nabla_{\perp}A_{1\|}\cdot \bm\nabla_{\perp}\widehat{A}_{1\|}
+\frac{\mu}{2 B} A_{1\|}\bm\nabla_{\perp}^2\widehat{A}_{1\|}+\frac{\mu}{2 B}\bm\nabla_{\perp}^2 A_{1\|}\widehat{A}_{1\|},
\label{delta_H_2_delta_A_1}
\end{eqnarray}
The contributions from $H_{\mathsf {2 polmix}}$  given by Eq.~(\ref{H_2polmix}) are:
\begin{eqnarray*}
\frac{\delta{\mathcal A}^{{\mathcal E}, ({\mathsf{Vl}})}_{\mathsf{polmix}}}{\delta\phi_1}\circ\widehat{\phi}_{1}+
\frac{\delta{\mathcal A}^{{\mathcal E}, ({\mathsf{Vl}})}_{\mathsf{polmix}}}{\delta A_{1\|}}\circ\widehat{A}_{1 \|}
\equiv
-\epsilon_\delta^2\int dV\ dW \ F_0\ \left(\frac{\delta H_{\mathsf{2 polmix}}}{\delta\phi_{1}}\circ\widehat{\phi}_{1}+
\frac{\delta H_{\mathsf{2 polmix}}}{\delta A_{1\|}}\circ\widehat{A}_{1 \|}\right).
\end{eqnarray*}
We compute the first variation using the functional derivative defined in Eq.~$(\ref{func_derivative})$. In order to separate the dynamical and the Noether contributions, we integrate the above expression by parts. Here we analyze the electrostatic term with the test function $\widehat{\phi}_1$. 
First, we remind that the phase space measure $dV\ dW\equiv B_{\|}^*\ d^3{\mathbf X}\  d p_{z}\ d\mu$ contains the guiding-center Jacobian $B_{\|}^*=B_{\|}^*({\mathbf X,p_{z},\mu})$, so special attention needs to be paid when using the Leibniz rule. The first contribution is given by:
\begin{eqnarray}
\nonumber
&&\frac{\delta{\mathcal A}^{{\mathcal E}, ({\mathsf{Vl}})}_{\mathsf{polmix}}}{\delta\phi_1}\circ\widehat{\phi}_1=
\epsilon_\delta^2 \int B_{\|}^*\ dV\ d p_{z}\ d\mu\ \frac{m c^2}{B^2}\ F_0 \left[\bm\nabla_{\perp}\phi_1-\frac{p_{z}}{m c}\bm\nabla_{\perp}A_{1\|}\right]\cdot \bm\nabla_{\perp}\widehat{\phi}_1,
\\
&=&
\epsilon_\delta^2\int dV\ dp_{z}\ d\mu\ \bm\nabla_{\perp}\cdot \left[B_{\|}^*\ \left(\frac{m c^2}{B^2} F_0\right) 
\left(\bm\nabla_{\perp}\phi_1-\frac{p_{z}}{m c}\bm\nabla_{\perp}A_{1\|}\right)\widehat{\phi}_1\right]
\label{Noether_polmix}
\\
&-&
\epsilon_\delta^2\int dV\ \bm\nabla_{\perp}\cdot \left[\int\ B_{\|}^*\ dp_{z}\ d\mu \left(\frac{m c^2}{B^2} F_0\right)
 \left(\bm\nabla_{\perp}\phi_1-\frac{p_{z}}{m c}\bm\nabla_{\perp}A_{1\|}\right)\right]\widehat{\phi}_1.
 \label{dyn_polmix}
\end{eqnarray}
Here Eq.~(\ref{Noether_polmix}) is the Noether contribution and  Eq.~(\ref{dyn_polmix}) is the dynamical part.
The parallel magnetic potential contribution with $\widehat{A}_{1\|}$ can be obtained in a similar way:
\begin{eqnarray}
\nonumber
&&\frac{\delta{\mathcal A}^{{\mathcal E}, ({\mathsf{Vl}})}_{\mathsf{polmix}}}{\delta A_{1 \|}}\circ{\widehat{A}_{1\|}}= \epsilon_\delta^2
\int B_{\|}^*\ dV \ d p_{z}\ d\mu\ \left(\frac{m c^2}{B^2}\ F_0\right) \frac{p_{z}}{m c}\left[-\bm\nabla_{\perp}\phi_1+\frac{p_{z}}{m c}\bm\nabla_{\perp}A_{1\|}\right]\cdot \bm\nabla_{\perp}\widehat{A}_{1\|}
\\
&=&
\epsilon_\delta^2\int dV\ dp_{z}\ d\mu\ \bm\nabla_{\perp}\cdot \left[B_{\|}^*\ \left(\frac{m c^2}{B^2} F_0\right) 
\frac{p_{z}}{m c}\left(-\bm\nabla_{\perp}\phi_1+\frac{p_{z}}{m c}\bm\nabla_{\perp}A_{1\|}\right)\widehat{A}_{1\|}\right]
\\
\label{Noether_polmix_magn}
&-&
\epsilon_\delta^2\int dV \ \bm\nabla_{\perp}\cdot \left[\int\ B_{\|}^*\ dp_{z}\ d\mu \left(\frac{m c^2}{B^2} F_0\right)
\frac{p_{z}}{m c} \left(-\bm\nabla_{\perp}\phi_1+\frac{p_{z}}{m c}\bm\nabla_{\perp}A_{1\|}\right)\right]\widehat{A}_{1\|}.
 \label{dyn_polmix_magn}
\end{eqnarray}
When introducing the equilibrium fluid density $n_0$ and the equilibrium current $\mathfrak{J}_0$:
\begin{eqnarray}
& &n_0\equiv \int B_{\|}^* dp_{z}\ d\mu\ F_0,\\
& &\mathfrak{J}_0\equiv c\ \int B_{\|}^* dp_{z}\ d\mu\ \frac{p_{z}}{m c}\ F_0,
\end{eqnarray}
we can write:
\begin{eqnarray*}
\frac{\delta{\mathcal A}^{{\mathcal E}, ({\mathsf{Vl}})}_{\mathsf{polmix}}}{\delta\phi_1}\circ\widehat{\phi}_1
&=&
\epsilon_\delta^2\int d^3{\mathbf X}\ \bm\nabla_{\perp}\cdot \left[\frac{m c^2}{B^2}\left(n_0\bm\nabla_{\perp}\phi_1-\frac{1}{c}\mathfrak{J}_0\bm\nabla_{\perp}A_{1\|}\right)\ \widehat{\phi}_1\right]
\\
&-&
\epsilon_\delta^2\int d^3{\mathbf X}\ \bm\nabla_{\perp}\cdot \left[\frac{m c^2}{B^2}\left(n_0\bm\nabla_{\perp}\phi_1-\frac{1}{c}\mathfrak{J}_0\bm\nabla_{\perp}A_{1\|}\right)\right] \widehat{\phi}_1.
\end{eqnarray*}
By defining the second moment of the equilibrium distribution function as
\begin{equation*}
\widetilde{\mathfrak J}_0\equiv c\ \int B_{\|}^* dp_{z}\ d\mu\ \left( \frac{p_{z}}{m c}\right)^2\ F_0,
\end{equation*}
we can write the electromagnetic part as:
\begin{eqnarray*}
\frac{\delta{\mathcal A}^{{\mathcal E}, ({\mathsf{Vl}})}_{\mathsf{polmix}}}{\delta A_{1\|}}\circ\widehat{A}_{1\|}
&=&
\epsilon_\delta^2\int d^3{\mathbf X}\ \bm\nabla_{\perp}\cdot \left[\frac{m c^2}{B^2}\left(-\frac{1}{c}\mathfrak{J}_0\bm\nabla_{\perp}\phi_1+\frac{1}{c^2}\widetilde{\mathfrak J}_0\bm\nabla_{\perp}A_{1\|}\right)\ \widehat{A}_{1\|}\right]
\\
&&-
\epsilon_\delta^2\int d^3{\mathbf X}\ \bm\nabla_{\perp}\cdot \left[\frac{m c^2}{B^2}\left(-\frac{1}{c}\mathfrak{J}_0\bm\nabla_{\perp}\phi_1+\frac{1}{c^2}\widetilde{\mathfrak J}_0\bm\nabla_{\perp}A_{1\|}\right)\right] \widehat{A}_{1\|}.
\end{eqnarray*}
Next we consider the dynamical and Noether contributions coming from the magnetic Vlasov part of the action, corresponding to the second order Hamiltonian ${H}_{\mathsf{2 polmag}}$, which contains the first and second order FLR corrections, given in Eq.~(\ref{H2_full_GK}):
\begin{eqnarray*}
&&\frac{\delta{\mathcal A}^{{\mathcal E}, ({\mathsf{Vl}})}_{\mathsf{polmag (1)}}}{\delta A_{1\|}}\circ\widehat{A}_{1\|}+
\frac{\delta{\mathcal A}^{{\mathcal E}, ({\mathsf{Vl}})}_{\mathsf{polmag (2)}}}{\delta A_{1\|}}\circ\widehat{A}_{1\|}
\nonumber
\\
&&\equiv
-\epsilon_\delta^2\int dV\ dW \ F_0\ \left(\frac{\delta H_{\mathsf{polmag(1)}}}{\delta A_{1\|}}\circ\widehat{A}_{1|\|}+
\frac{\delta H_{\mathsf{polmag(2)}}}{\delta A_{1\|}}\circ\widehat{A}_{1 \|}\right),
\end{eqnarray*}
where $\mathcal{A}^{{\mathcal E}, ({\mathsf{Vl}})}_{\mathsf{polmag (1)}}$ contains the first order FLR correction from $H_{\mathsf{2 polmag}}$ and 
$\mathcal{A}^{{\mathcal E}, ({\mathsf{Vl}})}_{\mathsf{polmag (2)}}$, the second order one.
The contribution from the variation of the first order FLR term is
\begin{eqnarray*}
\nonumber
\frac{\delta{\mathcal A}^{\mathcal E}_{\mathsf{polmag\ (1)}}}{\delta A_{1\|}}\circ\widehat{A}_{1\|}
&=&-\epsilon_{\delta}^2\ \frac{e^2}{m c^2}\int d V\ dW\ F_0 \ A_{1\|}\ \widehat{A}_{1\|}-
\epsilon_{\delta}^2\int dV\ dW\ \frac{\mu}{B}\ F_0\ \bm\nabla_{\perp} A_{1\|}\cdot\bm\nabla_{\perp} \widehat{A}_{1\|}
\\
&=&-\epsilon_{\delta}^2\ \frac{e^2}{m c^2}\int d V\ dW\ F_0 \ A_{1\|}\ \widehat{A}_{1\|}
+
\epsilon_{\delta}^2\int dV d\mu\ dp_{z} \bm\nabla_{\perp}\cdot\left[B_{\|}^*\ F_0\frac{\mu}{B}\bm\nabla_{\perp}A_{1\|}\right]\widehat{A}_{1\|}
\nonumber
\\
&-&
\epsilon_{\delta}^2\int dV\ d\mu\ dp_{z} \bm\nabla_{\perp}\cdot \left[B_{\|}^*\ F_0\frac{\mu}{B}\bm\nabla_{\perp}A_{1\|}\ \widehat{A}_{1\|}\right].
\nonumber
\end{eqnarray*}
The second order contribution is 
%
\begin{eqnarray*}
\frac{\delta{\mathcal A}^{{\mathcal E}, ({\mathsf{Vl}})}_{\mathsf{polmag\ (2)}}}{\delta A_{1\|}}\circ\widehat{A}_{1\|}
&=&
-\frac{\epsilon_{\delta}^2}{2}
\int\ dV\ dW\
\left(F_0\frac{\mu}{B}\right)\ \left(A_{1\|}\bm\nabla^2_{\perp} \widehat{A}_{1\|}+\bm\nabla^2_{\perp} A_{1\|}\ \widehat{A}_{1\|}\right).
\end{eqnarray*}
The second term of the right hand side of the above equation gives directly a dynamical contribution to the Amp\`ere's equation. By using two successive integration by parts, we rewrite the first term in order to obtain the Noether and dynamical contributions:
\begin{eqnarray}
&&-\int\ dV\ dW\
\left(F_0\frac{\mu}{B}\right)\ \left(A_{1\|}\bm\nabla^2_{\perp} \widehat{A}_{1\|}\right)\nonumber
\\
&& \qquad  =-
\int\ dV\ d\mu\ dp_{z}\left\{
\bm\nabla_{\perp}\cdot 
\left[
B_{\|}^*\left(F_0\frac{\mu}{B}\right)\ A_{1\|}\bm\nabla_{\perp} \widehat{A}_{1\|}
\right]
-
\bm\nabla_{\perp} \cdot 
\left[
B_{\|}^* F_0\frac{\mu}{B} \ A_{1\|} \right] \bm\nabla_{\perp}\widehat{A}_{1\|} \right\},\nonumber
\\
&&\qquad =-
\int\ dV\ d\mu\ dp_{z}
\bm\nabla_{\perp}\cdot 
\left[
\left(B_{\|}^* F_0\frac{\mu}{B}\right)\ A_{1\|}\bm\nabla_{\perp} \widehat{A}_{1\|}
\right]
\nonumber
\\
&&
\qquad +
\int\ dV\  d\mu\ dp_{z}
\bm\nabla_{\perp}\cdot 
\left[\bm\nabla_{\perp}
\left(B_{\|}^* F_0\frac{\mu}{B}\ A_{1\|}\right) \widehat{A}_{1\|}
\right]
-
\int\ dV\ d\mu\ dp_{z}
\bm\nabla_{\perp}^2
\left[
B_{\|}^* F_0\frac{\mu}{B}\ A_{1\|}
\right]
\widehat{A}_{1\|}.
\nonumber
\end{eqnarray}
The second order FLR contributions become:
\begin{eqnarray}
&&\frac{\delta{\mathcal A}^{{\mathcal E}, ({\mathsf{Vl}})}_{\mathsf{polmag\ (2)}}}{\delta A_{1\|}}\circ\widehat{A}_{1\|}
=-\frac{\epsilon_{\delta}^2}{2}
\int\ dV\  d\mu\ dp_{z}\
\bm\nabla_{\perp}\cdot 
\left[
\left(B_{\|}^*F_0\frac{\mu}{B}\right)\ A_{1\|}\bm\nabla_{\perp} \widehat{A}_{1\|}
\right] \nonumber 
\\
\nonumber
&&
+
\frac{\epsilon_{\delta}^2}{2}
\int\ dV\ d\mu\ dp_{z}\
\bm\nabla_{\perp}\cdot 
\left[\bm\nabla_{\perp}
\left(B_{\|}^*F_0\frac{\mu}{B}\ A_{1\|}\right) \widehat{A}_{1\|}
\right]
\\
&&
-\frac{\epsilon_{\delta}^2}{2}
\int\ dV\ d\mu\ dp_{z}\
\left[\bm\nabla_{\perp}^2
\left[
B_{\|}^*F_0\frac{\mu}{B}\ A_{1\|}
\right] +
\left(B_{\|}^*F_0\ \frac{\mu}{B}\right)\ \bm\nabla_{\perp}^2 A_{1\|}\right]  \widehat{A}_{1\|}.
\nonumber
\end{eqnarray}
From above, we notice that there are two Noether contributions and two dynamical ones.

Finally, the contribution from the magnetic part of the polarization can be written as:
\begin{eqnarray}
&&\frac{\delta{\mathcal A}^{{\mathcal E}, ({\mathsf{Vl}})}_{\mathsf{polmag}}}{\delta A_{1\|}}\circ\widehat{A}_{1\|}
=
\nonumber
\\
&-&\epsilon_{\delta}^2\frac{e^2}{m c^2}\int d V\ dW\ F_0 \ A_{1\|}\widehat{A}_{1\|}
+
\label{Vl_polmag1}
\epsilon_{\delta}^2\int\ dV\  d\mu\ dp_{z}\ \bm\nabla_{\perp}\cdot \left(B_{\|}^*F_0\frac{\mu}{B}\bm\nabla_{\perp}A_{1\|}\right)\widehat{A}_{1\|}\\
&-&\frac{\epsilon_{\delta}^2}{2}\int \ dV\ d\mu\ dp_{z}\ \bm\nabla^2_{\perp}\left(B_{\|}^*F_0\frac{\mu}{B}A_{1\|}\right)\ \widehat{A}_{1\|}
\label{Vl_polmag2}
-\frac{\epsilon_{\delta}^2}{2}
\int dV\ d\mu\ dp_{z}\ 
\left(B_{\|}^*F_0\frac{\mu}{B}\right)\bm\nabla_{\perp}^2A_{1\|}\widehat{A}_{1\|}
\\
\nonumber
&+&\frac{\epsilon_{\delta}^2}{2}\int dV\ d\mu\ dp_{z}\ \bm\nabla_{\perp}\cdot \left[\bm \nabla_{\perp}\left(B_{\|}^*F_0\frac{\mu}{B}A_{1\|}\right)\widehat{A}_{1\|}
-\left(B_{\|}^*F_0\frac{\mu}{B}A_{1\|}\right)\bm \nabla_{\perp}\widehat{A}_{1\|}
\right]
\\
&-&\epsilon_{\delta}^2\int dV\ d\mu\ dp_{z}\ \bm\nabla_{\perp}\cdot \left[B_{\|}^*F_0\frac{\mu}{B}\bm\nabla_{\perp}A_{1\|}\widehat{A}_{1\|}\right].
\nonumber
\end{eqnarray}

%


\begin{thebibliography}{10}

\bibitem{Bottino_Sonnendrucker}
A.~{Bottino} and E.~{Sonnendrucker}.
\newblock {Monte Carlo Particle-In-Cell methods for the simulation of the
  Vlasov-Maxwell gyrokinetic equations}.
\newblock {\em Journal of Plasma Physics}, 81(5):435810501, 2015.

\bibitem{Bottino_2011}
A.~{Bottino}, T.~{Vernay}, B.~D. {Scott}, S.~{Brunner}, and R.~{Hatzky}.
\newblock {Global simulations of tokamak microturbulence: finite-beta effects
  and collisions}.
\newblock {\em {Plasma Physics and Controlled Fusion}}, 53(12):124027, 2011.

\bibitem{brizard_prl_2000}
A.~J. {Brizard}.
\newblock {New Variational Principle for the Vlasov-Maxwell Equations}.
\newblock {\em Physical Review Letters}, 84(25):5768, 2000.

\bibitem{brizard_2010}
A.~J. {Brizard}.
\newblock {Exact energy conservation laws for full and truncated nonlinear
  gyrokinetic equations}.
\newblock {\em Physics of Plasmas}, 17(4):042303, 2010.

\bibitem{Brizard_Hahm}
A.~J. {Brizard} and T.~S. {Hahm}.
\newblock {Foundations of nonlinear gyrokinetic theory}.
\newblock {\em Reviews of Modern Physics}, 79(2):421--468, 2007.

\bibitem{Brizard_Tronko}
A.~J. {Brizard} and N.~{Tronko}.
\newblock {Exact momentum conservation laws for the gyrokinetic Vlasov-Poisson
  equations}.
\newblock {\em Physics of Plasmas}, 18(8):082307, 2011.

\bibitem{Cendra_Holm_Hoyle_Marsden_1998}
H.~{Cendra}, D.~D. {Holm}, M.~J.~W. {Hoyle}, and J.~E. {Marsden}.
\newblock {The Maxwell-Vlasov equations in Euler-Poincare form}.
\newblock {\em {Journal of Mathematical Physics}}, 1(3138):3138--3157, 1998.

\bibitem{Frieman_Chen_1982}
E.~A. {Frieman} and L.~{Chen}.
\newblock {Nonlinear gyrokinetic equations for low-frequency electromagnetic
  waves in general plasma equilibria}.
\newblock {\em Physics of Fluids}, 25(3):502--508, 1982.

\bibitem{Goerler_2011}
T.~{Goerler}, X.~{Lapillonne}, S.~{Brunner}, {Dannert} T., F.~{Jenko},
  F.~{Merz}, and D.~{Told}.
\newblock {The global version of the gyrokinetic turbulence code GENE }.
\newblock {\em {Journal of Computational Physics}}, 230(18):7053 -- 7071, 2011.

\bibitem{Hatzky_2007}
R.~{Hatzky}, A.~{Koenies}, and A.~{Mishchenko}.
\newblock {Electromagnetic gyrokinetic PIC simulation with an adjustable
  control variates method}.
\newblock {\em {Journal of Computational Physics}}, 225(1):568 -- 590, 2007.

\bibitem{Holm_Marsden_Ratiu_1998}
D.~D. {Holm}, J.~E. {Marsden}, and T.~S. {Ratiu}.
\newblock {The Euler- €"Poincare Equations and Semidirect Products with
  Applications to Continuum Theories}.
\newblock {\em {Advances in Mathematics}}, 137(1):1--81, 1998.

\bibitem{Jenko_2000}
F.~{Jenko}, W.~{Dorland}, M.~{Kotschenreuther}, and B.~N. {Rogers}.
\newblock {Electron temperature gradient driven turbulence}.
\newblock {\em {Physics of plasmas}}, 7(5):1904--1910, 2000.

\bibitem{Jolliet_2007}
S.~{Jolliet}, A.~{Bottino}, P.~{Angelino}, R.~{Hatzky}, T.~M. {Tran}, B.~F.
  {Mcmillan}, O.~{Sauter}, K.~{Appert}, Y.~{Idomura}, and L.~{Villard}.
\newblock {A global collisionless PIC code in magnetic coordinates}.
\newblock {\em Computer Physics Communications}, 177(5):409--425, 2007.

\bibitem{Lee_1986}
W.~W. {Lee}.
\newblock {Gyrokinetic particle simulation model}.
\newblock {\em {Journal of Computational Physics}}, 72(1):243 -- 269, 1987.

\bibitem{Littlejohn_1983}
R.~G. {Littlejohn}.
\newblock {Variational principles of guiding centre motion}.
\newblock {\em {Journal of Plasma Physics}}, 29(FEB):111--125, 1983.

\bibitem{Low_1958}
F.~E. {Low}.
\newblock {A Lagrangian formulation of the Boltzmann-Vlasov equations for
  plasmas}.
\newblock {\em {Proc.~R.~Soc.~London Series}}, 248(1253):282--287, 1958.

\bibitem{McMillan2012}
B.~F. {McMillan}, P.~{Hill}, S.~{Jolliet}, T.~{Vernay}, and L.~{Villard}.
\newblock {Gyrokinetic transport relations for gyroscale turbulence}.
\newblock {\em {Journal of Physics Conference series}}, 401:012014, 2012.

\bibitem{Scott_Smirnov}
B.~{Scott} and J.~{Smirnov}.
\newblock {Energetic consistency and momentum conservation in the gyrokinetic
  description of tokamak plasmas}.
\newblock {\em Physics of Plasmas}, 17(11):112302, 2010.

\bibitem{Squire_Qin_2013}
J.~{Squire}, H.~{Qin}, W.~M. {Tang}, and C.~{Chandre}.
\newblock {The Hamiltonian structure and Euler-Poincare formulation of the
  Vlasov-Maxwell and gyrokinetic systems}.
\newblock {\em Physics of Plasmas}, 20(2):022501, 2013.

\bibitem{Sugama_2000}
H.~{Sugama}.
\newblock {Gyrokinetic field theory}.
\newblock {\em Physics of Plasmas}, 7(2):466--480, 2000.

\bibitem{Tronko_Brizard_2015}
N.~{Tronko} and A.~J. {Brizard}.
\newblock {Lagrangian and Hamiltonian constraint for guiding-center Hamiltonian
  theories}.
\newblock {\em Physics of Plasmas}, 22(11):112507, 2015.

\end{thebibliography}

\end{document}